\documentclass[12pt]{amsart}
\usepackage{graphicx} 
\usepackage{amssymb}
\usepackage{amsmath}
\usepackage{amsthm}
\usepackage{ulem}
\usepackage{tikz}

\usepackage{color}
\definecolor{darkgreen}{rgb}{0,0.5,0}
\usepackage[colorlinks=true, linkcolor=darkgreen, urlcolor=blue, citecolor=red]{hyperref}
\usepackage{cleveref}
\usepackage{autonum}
\usepackage{tikz}
\usepackage{tikz-cd}

\newtheorem{proposition}{Proposition}
\newtheorem{lemma}{Lemma}
\newtheorem{theorem}{Theorem}
\newtheorem{definition}{Definition}
\newtheorem{remark}{Remark}
\newtheorem{corollary}{Corollary}

\begin{document}

\title[]{Low-dimensional tori in Calogero--Moser--Sutherland systems}

\author{Andrii Liashyk}

\address{A.L.:  Beijing Institute for Mathematical Sciences and Applications, Beijing, China}
\email{liashyk@bimsa.cn}

\author{Guorui Ma}

\address{G.M.:  Shanghai Institute for Mathematics and Interdisciplinary Sciences, Shanghai, China}
\email{maguorui@simis.cn}

\author{Nicolai Reshetikhin}

\address{N.R.: Department of Mathematics, Yau Center for Mathematical Sciences, Tsinghua University, Beijing, China;
Beijing Institute for Mathematical Sciences and Applications, Beijing, China;
St. Petersburg State University, St. Petersburg, Russia;
Department of Mathematics, UC Berkeley, USA}
\email{reshetik@math.berkeley.edu}

\author{Ivan Sechin}

\address{I.S.: Beijing Institute for Mathematical Sciences and Applications, Beijing, China}
\email{sechin@bimsa.cn}
\maketitle

\begin{abstract}
The main result of this paper is an explicit description of the stratification of the phase space of Calogero--Moser--Sutherland (CMS) integrable systems corresponding to Lie groups $SU(n)$. The phase space decomposes into symplectic strata of dimensions $2s$, where $s = 0, 1, \ldots, n - 1$. On each stratum of the positive dimension, we construct natural action-angle coordinates and compute the symplectic form explicitly, showing that every stratum is symplectomorphic to $\mathbb{R}_{> 0}^s \times \mathbb{T}^s$. The zero-dimensional stratum corresponds to the equilibrium point of the multi-time CMS dynamics.
\end{abstract}

\tableofcontents

\section{Introduction}

Recall that the Calogero--Moser--Sutherland (CMS) system \cite{Ca71, Mo75, Su71} is a model describing $n$ interacting particles on a circle with Hamiltonian
\begin{equation}
    \mathcal{H} = \frac{1}{2} \sum_{k = 1}^{n} p_k^2 + 
        \frac{1}{8} \sum_{\substack{k, l = 1 \\ k \ne l}}^{n} \frac{c^2}{\sin^2\left(\frac{q_k - q_l}{2}\right)}.
\end{equation}
Here $c^2$ is the coupling constant of the model, and throughout this paper we assume $c > 0$.

A natural geometric interpretation of the CMS system is provided by the Kazhdan--Kostant--Sternberg Hamiltonian reduction \cite{KKS78}. A brief review of this construction is given in Section~\ref{sec: CMS-system-and-reduction}. In this setting, the CMS phase space $\mathcal{S}(\mathcal{O})$ arises as a symplectic quotient of the cotangent bundle $T^* SU(n)$ equipped with its canonical symplectic structure, with respect to the adjoint action of $SU(n)$ at a value of the momentum map lying on the minimal nontrivial coadjoint orbit $\mathcal{O}$.

After performing the reduction, one can identify the reduced phase space $\mathcal{S}(\mathcal{O})$ with the quotient
\begin{equation}
    \mathcal{S}(\mathcal{O}) = T^* H_{\mathrm{reg}} / S_n.
\end{equation}
Here $H_{\mathrm{reg}} \subset SU(n)$ denotes the regular part of the Cartan subgroup of $SU(n)$, consisting of diagonal unitary matrices with pairwise distinct eigenvalues and unit determinant. 

The cotangent bundle $T^* H_{\mathrm{reg}}$ admits natural coordinates
\begin{equation}
    \big(p_k, \gamma_k = e^{i q_k} \big)_{k = 1}^n, \quad
        p_k \in \mathbb{R}, \quad 
        |\gamma_k| = 1,
\end{equation}
subject to the constraints
\begin{equation}
    \sum_{k = 1}^n p_k = 0, \quad
        \prod_{k = 1}^n \gamma_k = 1,
\end{equation}
together with the regularity condition $\gamma_k \ne \gamma_l$ for $k \ne l$. The symmetric group $S_n$ acts by simultaneous permutations of eigenvalues $\gamma_k$ and their conjugate momenta $p_k$. 

Since all eigenvalues are pairwise distinct on $H_{\mathrm{reg}}$, this $S_n$-action is free. Therefore, the quotient $\mathcal{S}(\mathcal{O}) = T^* H_{\mathrm{reg}} / S_n$ is a smooth manifold of dimension $2n - 2$.

The CMS system is a Liouville integrable system with $(n - 1)$ independent Poisson-commuting integrals
\begin{equation} \label{eq: Hams}
    \mathcal{H}_k = \frac{1}{k} \mathrm{Tr}(x^k), \quad k = 2, \ldots, n,
\end{equation}
where $x$ is the Lax matrix of the CMS system, expressed in terms of the phase space variables $p$ and $q$. For its explicit form, see Section~\ref{sec: Lax-matrix-and-Hamiltonians}.

The Poisson commutativity of the Hamiltonians $\mathcal{H}_k, \ k = 2, \ldots, n$ is a direct consequence of the Hamiltonian reduction. The quadratic Hamiltonian $\mathcal{H}_2$ reproduces the Calogero--Moser--Sutherland Hamiltonian $\mathcal{H}$.

In Section~\ref{sec: lagrangian-projection-and-stratification}, we give an explicit and uniform description of the stratification of the CMS phase space $\mathcal{S}(\mathcal{O})$. This is the main result of this paper. 

The Poisson-commuting Hamiltonians $\mathcal{H}_2, \ldots, \mathcal{H}_n$ generate a Hamiltonian action of the torus $\mathbb{T}^{n - 1}$ on $\mathcal{S}(\mathcal{O})$. The corresponding torus orbits are compact, connected, and, over generic points, are Lagrangian submanifolds of $\mathcal{S}(\mathcal{O})$. Thus, the map $\pi \colon \mathcal{S}(\mathcal{O}) \to \mathcal{B}(\mathcal{O})$, given by the joint values of the Hamiltonians, defines a Lagrangian fibration. The fibers of $\pi$ are the $\mathbb{T}^{n - 1}$-orbit of this Hamiltonian action. Since the Hamiltonians are symmetric functions of the eigenvalues of the Lax matrix $x$, the base $\mathcal{B}(\mathcal{O})$ can be described in terms of its eigenvalues $x_1, \ldots, x_n$ modulo permutations. See Section~\ref{sec:diag_x} for details.

In Section~\ref{sec: base-lagrangian-projection}, we prove that the image of the Lagrangian projection $\mathcal{B}(\mathcal{O}) = \mathrm{Im}(\pi)$ is a shifted principal Weyl chamber for $\mathfrak{su}(n)$, namely,
\begin{equation}
    \mathcal{B}(\mathcal{O}) = 
        \Big\{(x_1, \ldots, x_n) \, \Big| \, 
            x_{k - 1} - x_k \geq c, \ \text{for } k = 2, \ldots, n; \text{ and }
            \sum_{k = 1}^n x_k = 0 \Big\}.
\end{equation}

Our paper gives a concrete example of the general theory of symplectic toric manifolds. Atiyah \cite{At82} and Guillemin--Sternberg \cite{GS82} proved that for a Hamiltonian torus action on a \textit{compact} connected symplectic manifold, the image of the momentum map, interpreted as the base of the associated Lagrangian fibration, is a convex polytope whose vertices are the momentum map images of the fixed points. Delzant \cite{De88} showed that in this case the momentum polytope satisfies additional conditions and gave a complete classification of such polytopes. In the noncompact case, the image of the momentum map remains convex but is, in general, unbounded, and the role of Delzant polytopes is played by suitable convex polyhedra, see, e.g, Lerman--Tolman \cite{LT97} and Karshon--Lerman \cite{KL15}. The present paper provides an explicit example of this noncompact situation, describing a torus action on a noncompact symplectic manifold for which the momentum polytope becomes a cone.

The space $\mathcal{B}(\mathcal{O})$ is stratified:
\begin{equation}
    \mathcal{B}(\mathcal{O}) = \bigsqcup_{\{\mathbf{j}\}} \mathcal{B}(\mathcal{O})_{\{\mathbf{j}\}},
\end{equation}
where the disjoint union is taken over all subsets $\{\mathbf{j}\} \subseteq \{2, \ldots, n\}$. 

For a subset $\{\mathbf{j}\} = \{j_1, \ldots, j_s\}$, the corresponding component $\mathcal{B}(\mathcal{O})_{\{\mathbf{j}\}}$ is characterized by the conditions $x_{k - 1} - x_k > c$ if $k \in \{\mathbf{j}\}$ and $x_{k - 1} - x_k = c$ otherwise. When $s > 0$, $\mathcal{B}(\mathcal{O})_{\{\mathbf{j}\}}$ is naturally isomorphic to $\mathbb{R}_{> 0}^s$ for positive $s$, and it reduces to a single point for $s = 0$. 

The maximum stratum $\mathcal{B}(\mathcal{O})_{\{2,\dots, n\}}$ has the maximal dimension $n - 1$, which equals half of the dimension of the phase space. It coincides with the interior of the shifted principal Weyl chamber, where $x_{k - 1} - x_k > c$ for all $k = 2, \dots, n$. All remaining components appear as boundary faces and low-dimensional walls.

In Section~\ref{sec: induced-stratification-phase-space}, we show how the decomposition of the base $\mathcal{B}(\mathcal{O})$ induces the corresponding stratification of the reduced phase space $\mathcal{S}(\mathcal{O})$:
\begin{equation}
  \mathcal{S}(\mathcal{O}) = \bigsqcup_{\{\mathbf{j}\}} \mathcal{S}(\mathcal{O})_{\{\mathbf{j}\}},
\end{equation}
where each component is given by the preimage $\mathcal{S}(\mathcal{O})_{\{\mathbf{j}\}} = \pi^{-1}(\mathcal{B}(\mathcal{O})_{\{\mathbf{j}\}})$. We prove that each stratum of positive dimension is isomorphic to $\mathbb{R}_{> 0}^s \times \mathbb{T}^s$, where the torus factor corresponds to the Liouville torus of the CMS system.

\begin{figure}[ht]\label{BSU3}
\centering
\begin{tikzpicture}
    \draw[->, thick] (-5,-3) -- (1,-3) node[right] {$x_2 - x_3$}; 
    \draw[->, thick] (-4,-4) -- (-4,2) node[above] {$x_1 -x_2$}; 

    \fill[gray!15] (-3,-2) -- (-3,2) -- (1,2) -- (1,-2) -- cycle;
    
    \draw[black, thick] (-3,-4) -- (-3,2); 
    \fill[black] (-3, -3) circle (1pt) node[below left] {c}; 
    
    \draw[black, thick] (-5,-2) -- (1,-2); 
    \fill[black] (-4, -2) circle (1pt) node[below left] {c}; 


    \fill[black] (0, 0) circle (2pt) node[right] {$2$-dim}; 
    \fill[black] (-3, -0.5) circle (2pt) node[right] {$1$-dim}; 
    \fill[black] (-1, -2) circle (2pt) node[above right] {$1$-dim}; 
    \fill[black] (-3, -2) circle (2pt) node[below right] {$0$-dim}; 
    
\end{tikzpicture}
  \caption{The shifted principal Weyl chamber  for $G = SU(3)$ and dimensions of Liouville tori over the corresponding strata. }
\end{figure}

The maximal stratum $\mathcal{S}(\mathcal{O})_{\{2,\dots, n\}}$ is subject to the Arnold--Liouville Theorem and its structure is well-known, see, for example, \cite{Ru95}. Action-angle variables on this stratum are described in Proposition \ref{prop: ndim Omega}. Action-angle variables on low-dimensional strata are described in Proposition \ref{prop: sdim Omega}.

The minimal stratum $\mathcal{S}(\mathcal{O})_{\varnothing}$ is the fixed point of the multi-time Hamiltonian flow generated by Poisson commuting Hamiltonians $\mathcal{H}_k$ (see \cite[eq.~(5.11)]{Ru95} and \cite[eq.~(3.29)--(3.30)]{FA10}). In \cite[after eq.~(5.12)]{Ru95}, Ruijisenaars predicted that the dynamics contains all tori $\mathbb{T}^s,\, s = 0, \ldots, n - 1$. In this paper, we explicitly describe them.

In Section~\ref{sec: action-angle-variables-on-strata}, we construct explicit global coordinates $(y_{j_a}, \theta_{j_a})$, $a = 1, \ldots, s$, for each stratum. Here, the variables $y_{j_a}$ parametrize the base of the Lagrangian fibration, while $\theta_{j_a}$ are angular coordinates on the fibers. In these coordinates, the symplectic form on $\mathcal{S}(\mathcal{O})_{\{\mathbf{j}\}}$ takes the Darboux form
\begin{equation}
    \Omega_{\mathcal{S}(\mathcal{O})_{\{\mathbf{j}\}}} = 
        - \sum_{a = 1}^s \mathrm{d} y_{j_a} \wedge \mathrm{d} \theta_{j_a}, 
\end{equation}
so that $(y_{j_a}, \theta_{j_a})$ define action-angle variables on the stratum. This provides a complete description of the Lioville tori on lower-dimensional strata.

In Section~\ref{sec: multi-time-dynamics}, we analyze the multi-time CMS dynamics generated by the Poisson-commuting Hamiltonians $\mathcal{H}_2, \ldots, \mathcal{H}_n$. We show that these flows descend to each stratum of the reduced phase space and become linear in angle variables $\theta_{j_a}$. On lower-dimensional strata, the originally independent Hamiltonians become functionally dependent, reducing the number of angle variables. This phenomenon is illustrated in detail on the simplest nontrivial example of a two-dimensional stratum ($s = 1$), where all Hamiltonian flows act along a single angular direction, in Section~\ref{sec: motion-on-two-dim-stratum}.

This paper is closely related to the work \cite{LRS25} on hybrid integrable systems and to the upcoming paper \cite{CJRX25} on stratified superintegrable systems. 

Similar stratification exists in many other models, for example, in the elliptic Calogero--Moser system and integrable systems on moduli spaces of flat connections. We will address this in a separate publication. We also do not discuss the algebro-geometric setting, which is a very interesting subject and should be discussed separately.  

\section*{Acknowledgment}
We are grateful to L. Feher for stimulating discussions and pointing out reference \cite{FA10}, to K.~Jiang, Z.~Chen, and H.~Xiao for many discussions on stratified integrable systems.

The work of A.~L. was supported by the Beijing Natural Science Foundation (IS24006) and the Beijing Talent Program.
The work of N.~R. was supported by the Collaboration Grant "Categorical Symmetries" from
the Simons Foundation, by the Changjiang fund, and by the project 075-15-2024-631 funded
by the Ministry of Science and Higher Education of the Russian Federation.

\section{The Calogero--Moser--Sutherland system} \label{sec: CMS-system-and-reduction}

\subsection{Hamiltonian reduction along the coadjoint orbit}\label{CMS-n}
Here we review some facts about the CMS system for $SU(n)$.
This section is included for completeness of the exposition. It is a well-known Kazhdan--Kostant--Sternberg construction \cite{KKS78} of the Calogero--Moser--Sutherland system by the Hamiltonian reduction of $T^* G$ for $G = SU(n)$ with respect to the adjoint action. 

Denote by $\mathfrak{su}(n)$ a Lie algebra of $SU(n)$. We treat elements of $\mathfrak{su}(n)$ as traceless $n \times n$ anti-Hermitian matrices, and elements of $SU(n)$ as $n\times n$ unitary matrices with determinant $1$. We use the $\mathrm{Tr}$ form to identify the elements of $\mathfrak{su}(n)$ with the elements of its dual space $\mathfrak{su}(n)^*$. For $\mathfrak{su}(n)$ this means that both $\mathfrak{su}(n)$ and $\mathfrak{su}(n)^*$ are represented by traceless anti-Hermitian $n \times n$ matrices.

We trivialize the cotangent bundle by the right translations $T^* SU(n) \simeq \mathfrak{su}(n)^* \times SU(n)$.
We will parametrize this space by a unitary matrix $g \in SU(n)$ with $\det(g) = 1$ and a Hermitian traceless matrix $x$ such that $i \, x \in \mathfrak{su}(n)^*$ 
\begin{equation}
    T^* SU(n) \simeq
        \{ (x, g) \mid x = x^\dagger, \ \mathrm{Tr}(x) = 0, \ g \in SU(n) \}.
\end{equation}
The canonical symplectic form on $T^* SU(n)$ in these coordinates is given by
\begin{equation} \label{eq: symplectic_form_T^*G}
    \Omega = i \, \mathrm{d} \, \mathrm{Tr}(x\, \mathrm{d} g \, g^{-1}).
\end{equation}
The adjoint action of $SU(n)$ on itself $g \mapsto h g h^{-1}$ could be lifted to the Hamiltonian action of $SU(n)$ on $T^* SU(n)$: $g \mapsto h g h^{-1}, \ x \mapsto h x h^{-1}$ with the momentum map
\begin{equation} \label{eq: phase_space_equation}
    \mu \colon T^* SU(n) \to \mathfrak{su}(n)^*, \quad
    \mu(x, g) = i (x - g^{-1} x g). 
\end{equation}

The Hamiltonian reduction for the adjoint action of $SU(n)$ on $T^*SU(n)$ along the coadjoint orbit $\mathcal{O}\subset \mathfrak{su}(n)^*$ produces the symplectic leaf of the Poisson space $T^*SU(n)/SU(n)$ which is 
\begin{equation}\label{slCM}
    \mathcal{S}(\mathcal{O}) = \mu^{-1}(\mathcal{O})/SU(n) =
        \{(x, g) \mid x = x^\dagger, \ \mathrm{Tr}(x) = 0, \ g \in SU(n), \ \mu(x, g) \in \mathcal{O}\} / SU(n).
\end{equation}
The symplectic form $\Omega_{\mathcal{S}(\mathcal{O})}$ on the reduced phase space is uniquely determined by the relation $\mathrm{pr}^* \Omega_{\mathcal{S}(\mathcal{O})} = \iota^* \Omega$, where $\iota \colon \mu^{-1}(\mathcal{O}) \hookrightarrow T^* SU(n)$ is the inclusion and $\mathrm{pr} \colon \mu^{-1}(\mathcal{O}) \to \mu^{-1}(\mathcal{O})/SU(n) \simeq \mathcal{S}(\mathcal{O})$ is the quotient projection
\begin{equation}
    \begin{tikzcd}
        \mu^{-1}(\mathcal{O}) \arrow["\mathrm{pr}", d]
        \arrow["\iota", r]
        &  T^*SU(n) \\
        \mathcal{S}(\mathcal{O})&
    \end{tikzcd}
\end{equation}

\subsection{Minimal nontrivial coadjoint orbit and CMS phase space}

To get the phase space of the CMS model, we choose $\mathcal{O}$ to be a minimal nontrivial coadjoint orbit of $SU(n)$, i.e., a coadjoint orbit of the smallest positive dimension. Such orbits correspond to elements of $\mathfrak{su}(n)^*$ whose eigenvalue multiplicities are $(1, n - 1)$. 

The dimension of $\mathcal{O}$ is $2n - 2$. Its elements can be represented by traceless anti-Hermitian matrices of the form $i \big( \psi \psi^\dagger - c \cdot 1 \big)$, where $\psi \in \mathbb{C}^n$ satisfies $\psi^\dagger \psi = nc$. Multiplication of $\psi$ by a phase factor $e^{i \theta}$ leaves the matrix $\psi \psi^\dagger$, and hence the corresponding element of $\mathcal{O}$, unchanged. Thus, the orbit $\mathcal{O}$ can be equivalently described as
\begin{equation}
    \mathcal{O} \simeq \left\{
            \psi \in \mathbb{C}^n \mid \psi^\dagger \psi = n c
        \right\} \big/ U(1),
\end{equation}
where $U(1)$ acts by phase rotations $\psi \mapsto e^{i \theta} \psi$.
As a smooth manifold, $\mathcal{O} \simeq \mathbb{CP}^{n - 1}$.

It is clear that elements of $\mathcal{O}$ have eigenvalues $i (n - 1)c$ (with multiplicity $1$) and $-i c$ (with multiplicity $(n - 1)$).

For this choice of the coadjoint orbit $\mathcal{O}$, the reduced phase space $\mathcal{S}(\mathcal{O})$ can be described explicitly as follows. It consists of equivalence classes of triples $(x, g, \psi)$, where $x$ is a traceless Hermitian matrix, $g \in SU(n)$, and $\psi \in \mathbb{C}^n$, subject to the momentum map equation
\begin{equation} \label{eq: momentum_map_equation}
    x - g^{-1} x g = \psi \psi^\dagger - c \cdot 1.
\end{equation}
More precisely,
\begin{equation} \label{eq: explicit_form_of_S(O)}
    \mathcal{S}(\mathcal{O}) = \{ (x, g, \psi) \mid 
        x^\dagger = x, \ \mathrm{Tr}(x) = 0, \ g \in SU(n), \ \psi^\dagger \psi = n c, \ 
        x - g^{-1} x g = \psi \psi^\dagger - c \cdot 1 \} \big/ U(n),
\end{equation}
with the $U(n)$-action
\begin{equation}
    x \mapsto h x h^{-1}, \quad
    g \mapsto h g h^{-1}, \quad
    \psi \mapsto h \psi, \quad
    h \in U(n).
\end{equation}
Although $\mathcal{S}(\mathcal{O})$ is defined as $\mu^{-1}(\mathcal{O})/SU(n)$, the explicit parametrization above naturally leads to a quotient by $U(n)$, since the description of $\mathcal{O}$ involves the additional $U(1)$ phase rotation acting on $\psi$. This $U(1)$-action acts trivially on $x$ and $g$, and then can be combined with the $SU(n)$-action, resulting in an effective $U(n)$-action on the triples $(x, g, \psi)$.

\subsection{The diagonalization of $g$ and natural coordinates}\label{sec:diagx}

Let us describe the reduced phase space $\mathcal{S}(\mathcal{O})$ in the gauge where the matrix $g$ is diagonal. Since any unitary matrix can be diagonalized by a unitary transformation, the $U(n)$-action allows us to choose a representative of the form
\begin{equation}\label{eq: gdiag}
    g = \gamma = \mathrm{diag}(\gamma_1, \ldots, \gamma_n), \quad 
        |\gamma_k| = 1, \text{ for } k = 1, \ldots, n, \quad 
            \prod_{l = 1}^n \gamma_l = 1.
\end{equation}

In this gauge, the momentum map equation \eqref{eq: momentum_map_equation} takes the component form
\begin{align}
    \label{eq: MME_g_gauge_diagonal}
    0 = |\psi_k|^2 - c, \qquad \qquad  \qquad  &k = l, \\
    \label{eq: MME_g_gauge_offdiagonal}
    x_{kl} (1 - \gamma_k^{-1} \gamma_l) = \psi_k \bar{\psi_l}, \qquad & k \ne l.
\end{align}

The diagonal equation \eqref{eq: MME_g_gauge_diagonal} implies that $|\psi_k|^2 = c$ with $c > 0$ for all $k = 1, \ldots, n$, and thus all components $\psi_k$ are nonzero. If $\gamma_k = \gamma_l$ for some $k \ne l$, the off-diagonal equation \eqref{eq: MME_g_gauge_offdiagonal} reduces to $\psi_k \bar{\psi_l} = 0$, which contradicts this condition. Therefore, all entries of the diagonal matrix $g$ are distinct: $\gamma_k \ne \gamma_l$ for $k \ne l$. Together with $\prod_{l = 1}^n \gamma_l = 1$, it shows that $g \in H_{\mathrm{reg}}$.

After fixing the gauge so that $g$ is diagonal with pairwise distinct eigenvalues, the remaining gauge symmetry is given by the normalizer $N_{U(n)}(U(1)^{\times n})$ of the maximal torus $U(1)^{\times n} \subset U(n)$. The torus $U(1)^{\times n}$ is a normal subgroup of its normalizer, and the quotient $N_{U(n)}(U(1)^{\times n})\big/U(1)^{\times n}$ is naturally identified with the Weyl group $S_n$. 

Using the $U(1)^{\times n}$-symmetry, we can fix phases of $\psi_k, \ k = 1, \ldots, n$ and choose a representative with
\begin{equation} \label{eq: psi_with_fixed_phases}
    \psi_k = \sqrt{c}, \quad k = 1, \ldots, n.
\end{equation}
Substituting \eqref{eq: psi_with_fixed_phases} into the \eqref{eq: MME_g_gauge_offdiagonal}, we obtain
\begin{equation}
    x_{kl} = \frac{c}{1 - \gamma_k^{-1} \gamma_l}, \quad k \ne l.
\end{equation}
As a result, in this gauge, the off-diagonal entries of $x$ are completely determined by the eigenvalues of $g$, while the diagonal elements $x_{kk}$ remain unconstrained. Let us introduce the variables $p_k = x_{kk}, \ k = 1, \ldots, n$. Since the matrix $x$ is Hermitian and traceless, $p_k$ are real and satisfy the constraint $\sum_{l = 1}^n p_l = 0$. Accordingly, the diagonal matrix $i \, p$, where $p = \mathrm{diag}(p_1, \ldots, p_n)$, can be regarded as an element of $\mathfrak{h}^*$, the dual to the Cartan subalgebra $\mathfrak{h} \subset \mathfrak{su}(n)$. 

This shows that the reduced phase space can be identified with the quotient of $T^* H_{\mathrm{reg}} \simeq \mathfrak{h}^* \times H_{\mathrm{reg}}$ (with coordinates $(p_k, \gamma_k)$) by the residual action of the Weyl group $S_n$, which acts by simultaneous permutation of the pairs $(p_k, \gamma_k)$. Explicitly,
\begin{equation}
    \mathcal{S}(\mathcal{O}) = 
        \Big\{ (p, \gamma) \, \Big| \, 
            p_k \in \mathbb{R}, \ |\gamma_k| = 1 \text{ for } k = 1, \ldots, n, \ 
            \sum_{l = 1}^n p_l = 0, \ \prod_{l = 1}^n \gamma_l = 1 \Big\} \Big/ S_n 
                \simeq T^* H_{\mathrm{reg}}/S_n.
\end{equation}
Clearly, the Weyl group $S_n$ acts freely on $H_{\mathrm{reg}}$, and hence on $T^* H_{\mathrm{reg}}$. Therefore, the quotient $\mathcal{S}(\mathcal{O}) = T^* H_{\mathrm{reg}}/S_n$ is a smooth manifold of dimension $2n - 2$. 
Moreover, since $H_{\mathrm{reg}}$ is connected, the reduced phase space $\mathcal{S}(\mathcal{O})$ is connected as well.

The symplectic form on $\mathcal{S}(\mathcal{O})$ is induced from the canonical symplectic form on the cotangent bundle $T^* H_{\mathrm{reg}}$, which in coordinates $(p_k, \gamma_k)$ takes the form
\begin{equation}
    \Omega_{T^* H_{\mathrm{reg}}} = i \, \sum_{k = 1}^n \mathrm{d} p_k \wedge \gamma_k^{-1} \mathrm{d} \gamma_k,
\end{equation}
subject to the constraints $\sum_{l =  1}^n p_l = 0$ and $\prod_{l = 1}^n \gamma_l = 1$. 
This form is invariant under the $S_n$-action and therefore descends to a well-defined symplectic form $\Omega_{\mathcal{S}(\mathcal{O})}$ on the quotient.

\subsection{Lax matrix and Hamiltonians} \label{sec: Lax-matrix-and-Hamiltonians}

In this gauge, the explicit form of the matrix $x$ gives the Lax matrix of the CMS model
\begin{equation}\label{eq: Lax}
    x(p, \gamma) = \begin{pmatrix}
        p_1 & \frac{c}{1 - \gamma_1^{-1} \gamma_2} &
            \ldots & \frac{c}{1 - \gamma_1^{-1} \gamma_n} \\
        \frac{c}{1 - \gamma_2^{-1} \gamma_1} & p_2 & \ldots & \frac{c}{1 - \gamma_2^{-1} \gamma_n} \\
        \vdots & \vdots & \ddots & \vdots \\
        \frac{c}{1 - \gamma_n^{-1} \gamma_1} & \frac{c}{1 - \gamma_n^{-1} \gamma_2} & \ldots & p_n
    \end{pmatrix}.
\end{equation}

Consider the family of Poisson-commutative $SU(n)$-invariant functions on $T^* SU(n)$ defined by
\begin{equation} \label{eq: CMHams}
    \mathcal{H}_k(x, g) = \frac{1}{k} \mathrm{Tr}(x^k), \quad \ k=2,\dots, n.
\end{equation}
Since the functions $\mathcal{H}_k(x, g)$ are invariant under the $SU(n)$-action, their restrictions to $\mu^{-1}(\mathcal{O})$ descend to well-defined functions on the reduced phase space $\mathcal{S}(\mathcal{O})$. Moreover, the functions $\{\mathcal{H}_k\}_{k = 2}^n$ Poisson commute and are functionally independent on $\mathcal{S}(\mathcal{O})$. As $\dim \mathcal{S}(\mathcal{O}) = 2n - 2$, they form a complete set of integrals of motion and hence define a Liouville integrable system on $\mathcal{S}(\mathcal{O})$, which we refer to as the CMS system. The Hamiltonians $\mathcal{H}_k$ are called the CMS Hamiltonians.

For the first two Hamiltonians in coordinates $(p_k, q_k)$ with $\gamma_k = e^{i q_k}$, we have
\begin{equation}
   \mathcal{H}_2 = \frac{1}{2} \mathrm{Tr} \big(x^2 \big) =
    \frac{1}{2} \sum_{k = 1}^n p_k^2 +
    \frac{1}{2} \sum_{\substack{k, l = 1 \\ k \ne l}}^n
    \frac{c^2}{(1 - \gamma_l^{-1} \gamma_k)(1 - \gamma_k^{-1} \gamma_l)}
    = 
     \frac{1}{2} \sum_{k = 1}^n p_k^2 +
      \frac{1}{8} \sum_{\substack{k, l = 1 \\ k \ne l}}^n
            \frac{c^2}{\sin^2 ( \tfrac{q_k - q_l}{2})},
\end{equation}

\begin{equation}
    \mathcal{H}_3 = \frac{1}{3} \mathrm{Tr} \big(x^3 \big) =
        \frac{1}{3} \sum_{k = 1}^n p_k^3 +
        \sum_{\substack{k, l = 1 \\ k \ne l}}^n
            \frac{c^2 \, p_{k}}{(1 - \gamma_l^{-1} \gamma_k)(1 - \gamma_k^{-1} \gamma_l)}
            = 
        \frac{1}{3} \sum_{k = 1}^n p_k^3 +
       \frac{1}{4} \sum_{\substack{k, l = 1 \\ k \ne l}}^n
            \frac{c^2 \, p_{k}}{{\sin^2 ( \tfrac{q_k - q_l}{2})}}.     
\end{equation}

\section{Base of the Lagrangian projection for the CMS model and stratification of $\mathcal{S}(\mathcal{O})$} \label{sec: lagrangian-projection-and-stratification}

\subsection{CMS system as a Lagrangian fibration}\label{sec:diag_x}
Geometrically, any Liouville integrable system on a $2d$-dimensional symplectic manifold $\mathcal{M}$, given by $d$ independent Poisson-commuting functions, naturally gives rise to a Lagrangian fibration, i.e., a surjective map $\pi \colon \mathcal{M} \to \mathcal{B}$, such that a generic fiber is Lagrangian. For regular values of $\pi$, the fibers are smooth $d$-dimensional Lagrangian tori by the Arnold--Liouville theorem.

In the case of the CMS system, the base $\mathcal{B}$ of the Lagrangian projection can be described as the space of spectral data of the Lax matrix $x$, which can be naturally identified with the quotient $\mathfrak{h}^*/S_n \simeq \mathfrak{su}(n)^*/SU(n)$. Equivalently, we can write
\begin{equation}
    \mathcal{B} \simeq 
        \Big\{ (x_1, \ldots, x_n) \in \mathbb{R}^n \, \Big| \, \sum_{l = 1}^n x_l = 0 \Big\} \Big/ S_n.
\end{equation}
Choosing the principal Weyl chamber in $\mathfrak{h}^*$ provides a convenient set of representatives for this quotient, namely
\begin{equation}
    \mathcal{B} \simeq
        \Big\{ (x_1, \ldots, x_n) \in \mathbb{R}^n \, \Big| \,
        x_1 \ge x_2 \ge \ldots \ge x_n, \ \sum_{l = 1}^n x_l = 0 
        \Big\} \subset \mathbb{R}^{n - 1}.
\end{equation}

From the results of the previous section, the CMS Lagrangian projection is
\begin{equation}\label{LP}
    \pi \colon T^* H_{\mathrm{reg}} / S_n \to \mathfrak{h}^* / S_n.
\end{equation}
In the coordinates $(p_k, \gamma_k)$ on $T^* H_{\mathrm{reg}}$, the Lagrangian projection is given by sending a point $(p_k, \gamma_k)$ to the spectrum of the Lax matrix $x(p, \gamma)$ \eqref{eq: Lax}.

\subsection{Base of the Lagrangian projection} \label{sec: base-lagrangian-projection}
Here we describe the image $\mathcal{B}(\mathcal{O}) \subset \mathcal{B}$ of the Lagrangian projection (\ref{LP}). We show that it is a convex set, namely, a shifted principal Weyl chamber, which can be regarded as a stratified space. We also describe the induced stratification of the phase space $\mathcal{S}(\mathcal{O})$ by the preimages of walls of this Weyl chamber.

It is convenient to work with the reduced phase space $\mathcal{S}(\mathcal{O})$ \eqref{eq: explicit_form_of_S(O)} in a gauge where the Hermitian matrix $x$ is diagonalized by a unitary transformation and the eigenvalues are ordered 
\begin{equation} \label{eq: xdiag}
 x = \mathrm{diag} \left(x_1, \ldots, x_n\right), \quad
     x_1 \ge x_2 \ge \dots \ge x_n, \quad
     \sum_{l = 1}^n x_l = 0.
\end{equation}
In this gauge, the Lagrangian projection $\pi$ maps the equivalence class of a triple $[(x, g, \psi)] \in \mathcal{S}(\mathcal{O})$ to the diagonal matrix $x$, viewed as a point of the base $\mathcal{B}$.

In the following, we will use a standard fact from linear algebra.
\begin{lemma}\label{lem: rank1 det}
Let $D$ be a diagonal $n \times n$ matrix with entries $D_{ab} = D_a \delta_{ab}$ and let $P$ be a rank-one matrix of the form $P_{ab} = u_a v_b$. Then
\begin{equation} \label{eq: rank1 det}
    \det(D - P - z \cdot 1) = 
        \prod_{a = 1}^n (D_a - z) - 
            \sum_{b = 1}^n u_b v_b \prod_{\substack{a = 1 \\a \ne b}}^n (D_a - z).
\end{equation}
\end{lemma}

Define an auxiliary vector $\lambda = g \psi$.

\begin{theorem} \label{thr: generic strata} \footnote{The content of this theorem is analogous to Lemma 3.4 in \cite{FA10}.}
In the gauge where the Hermitian matrix $x$ is diagonal \eqref{eq: xdiag}, the solutions to the momentum map equation \eqref{eq: momentum_map_equation} admit the following description.

\begin{enumerate}
\item[(a)] The diagonal entries of $x$ satisfy
\begin{equation}
    x_{k - 1} - x_k \ge c, \quad k = 2, \ldots, n,
\end{equation}
and the components of $\psi$ and $\lambda$ are constrained by
\begin{equation}
    |\psi_{1}|^{2} > 0, \quad |\lambda_{n}|^{2} > 0,
\end{equation}
together with the obvious nonnegativity constraints
\begin{equation}
    |\psi_{k}|^{2} \ge 0, \quad |\lambda_{k-1}|^{2} \ge 0, \quad k = 2, \ldots, n.
\end{equation}

\item[(b)] For any $k = 2, \ldots, n$, the following conditions are equivalent:
\begin{equation}
     x_{k - 1} - x_{k} = c, \quad
     \psi_{k} = 0, \quad 
     \lambda_{k - 1} = 0.
\end{equation}

\item[(c)] For $k = 1, \ldots, n$, the components of $\psi$ and $\lambda$ are given explicitly by
\begin{equation}\label{eq: psi lam expl}
    |\psi_{k}|^2 = c \prod_{\substack{l = 1 \\ l \ne k}}^n \frac{x_k - x_l + c}{x_k - x_l}, \quad
    |\lambda_k|^2 = c \prod_{\substack{l = 1 \\ l \ne k}}^n \frac{x_l - x_k + c}{x_l - x_k}.
\end{equation}

\end{enumerate}
\end{theorem}

\begin{proof}

Assume first that $x$ has simple spectrum, i.e., $x_1 > x_2 > \ldots > x_n$.

Rewrite the momentum map equation \eqref{eq: momentum_map_equation} in the form
\begin{equation}\label{eq: moment map diag}
    g^{-1} x g = x + c \cdot 1 - \psi \psi^{\dagger}.
\end{equation}
Taking characteristic polynomials of both sides, we obtain
\begin{equation}\label{eq: char poly}
    \prod_{k = 1}^n (x_k - z) = 
        \prod_{k = 1}^n (x_k + c - z) - 
            \sum_{k = 1}^n |\psi_k|^2 \prod_{\substack{l = 1 \\ l \ne k}}^n (x_l + c - z),
 \end{equation}
where we used Lemma~\ref{lem: rank1 det} to compute the determinant on the right-hand side.

Evaluating \eqref{eq: char poly} at $z = x_k + c$ gives
\begin{equation}\label{eq: xpsi}
    c \prod_{\substack{l = 1 \\ l \ne k}}^n (x_k - x_l + c) =
        |\psi_k|^2 \prod_{\substack{l = 1 \\ l \ne k}}^n (x_k - x_l), 
\end{equation}
and hence for regular $x$ with distinct eigenvalues,
\begin{equation}
    |\psi_k|^2 = c \prod_{\substack{l = 1 \\ l \ne k}}^n \frac{x_k - x_l + c}{x_k - x_l}.
\end{equation}

In particular, the equation \eqref{eq: xpsi} for $k = 1$ implies that $|\psi_1|$ is strictly positive
\begin{equation}\label{eq: xpsi1}
    |\psi_1|^2 = c \prod_{l = 2}^n \frac{x_1 - x_l + c}{x_1 - x_l} > 0. 
\end{equation}

For $k = 2$, we rewrite \eqref{eq: xpsi} as
\begin{equation}
    |\psi_2|^2 = c \, \frac{x_2 - x_1 + c}{x_2 - x_1} \prod_{l = 3}^n \frac{x_2 - x_l + c}{x_2 - x_l} =
        c \, \frac{x_1 - x_2 - c}{x_1 - x_2} \prod_{l = 3}^n \frac{x_2 - x_l + c}{x_2 - x_l}.
\end{equation}
Since $x_2 > x_l$ for all $l > 2$, the product over $l > 2$ is strictly positive. Moreover, $x_1 - x_2 > 0$. Therefore, the nonnegativity condition $|\psi_2|^2 \ge 0$ implies $x_1 - x_2 - c \ge 0$. The bound $x_1 - x_2 = c$ is saturated if and only if $|\psi_2|^2 = 0$, equivalently $\psi_2 = 0$.

Continuing this reasoning for $k > 2$, we get $x_{k - 1} \ge x_k + c$ from $|\psi_k|^2 \ge 0$, and the equality $x_{k - 1} - x_k = c$ is achieved if and only if $\psi_k = 0$.

The momentum map equation \eqref{eq: momentum_map_equation} can be rewritten for $\lambda$ as
\begin{equation}\label{eq: moment map diag}
    g x g^{-1} = x  - c \cdot 1 + \lambda \lambda^{\dagger}. 
\end{equation}
By the analogous arguments, we get $|\lambda_n| > 0$ and that for $k = 2, \ldots, n$ the condition $x_{k - 1} - x_k = c$ is equivalent to $\lambda_{k - 1} = 0$.

We have described the solutions of the momentum map equation \eqref{eq: momentum_map_equation} for regular elements $x$ and derived the inequalities $x_{k - 1} - x_k \ge c$ for their eigenvalues. The case where the spectrum of $x$ is not simple can not occur due to the connectedness of $\mathcal{S}(\mathcal{O})$, see \cite[Lemma 3.4]{FA10}.

\end{proof}

As a direct consequence of the inequalities in Theorem~\ref{thr: generic strata}(a), the image $\mathcal{B}(\mathcal{O})$ of the Lagrangian projection can be described explicitly.
\begin{corollary} 
The image of the projection (\ref{LP}) is the convex set $\mathcal{B}(\mathcal{O})\subset \mathbb{R}^{n - 1}$ which can be realized as the shifted principal Weyl chamber for $SU(n)$
\begin{equation}\label{BO}
    \mathcal{B}(\mathcal{O}) = 
        \Big\{(x_1, \ldots, x_n) \, \Big| \, 
            x_{k - 1} - x_k \geq c, \ \text{for } k = 2, \ldots, n; \text{ and }
            \sum_{k = 1}^n x_k = 0 \Big\}.
\end{equation}
\end{corollary}

This space admits a natural stratification.

\begin{definition}
Let $s$ be an integer such that $0 \le s \le n - 1$. Let $\{\mathbf{j}\} = \{j_1,\dots, j_s\}$ be a subset of $\{2, \ldots, n\}$, such that $2 \le j_1 < \ldots < j_s \leq n$ (in particular, $\{\mathbf{j}\} = \varnothing$ in case if $s = 0$), and set $\overline{\{\mathbf{j}\}} = \{2, \ldots, n\} \setminus \{\mathbf{j}\}$. 

We define $\mathcal{B}(\mathcal{O})_{\{\mathbf{j}\}}$ as the subset of $\mathcal{B}(\mathcal{O})$, given by the conditions $x_{k - 1} - x_k = c$ for $k \in \overline{\{\mathbf{j}\}}$, and $x_{j_a - 1} - x_{j_a} > c$ for $a = 1, \ldots, s$.
\end{definition}

Clearly, as a smooth manifold $\mathcal{B}(\mathcal{O})_{\{\mathbf{j}\}} \simeq \mathbb{R}^{s}_{> 0}$ if $s \ge 1$, while $\mathcal{B}(\mathcal{O})_\varnothing$ is a single point.

\begin{corollary}
    The shifted principal Weyl chamber $\mathcal{B}(\mathcal{O})$ admits a stratification by submanifolds of dimensions $0 \le s \le n - 1$
\begin{equation}\label{eq: Bstat}
    \mathcal{B}(\mathcal{O}) = \bigsqcup_{\{\mathbf{j}\}} \mathcal{B}(\mathcal{O})_{\{\mathbf{j}\}},
\end{equation}
where the disjoint union is taken over all subsets $\{\mathbf{j}\} \subseteq \{2, \ldots, n\}$.
\end{corollary}

The example of such stratification for $G = SU(3)$ is illustrated in Figure \ref{BSU3} in the Introduction.

\subsection{Description of strata of $\mathcal{S}(\mathcal{O})$} \label{sec: induced-stratification-phase-space}

The stratification of $\mathcal{B}(\mathcal{O})$ induces a stratification of the reduced phase space $\mathcal{S}(\mathcal{O})$ by the preimages of $\mathcal{B}(\mathcal{O})_{\{\mathbf{j}\}}$ under the Lagrangian projection. In this section, we describe the resulting strata of $\mathcal{S}(\mathcal{O})$.

\begin{definition}
Let $s$ be an integer such that $0 \le s \le n - 1$. Let $\{\mathbf{j}\} = \{j_1,\dots, j_s\} \subseteq \{2, \ldots, n\}$ with $2 \le j_1 < \ldots < j_s \leq n$.

We define $\mathcal{S}(\mathcal{O})_{\{\mathbf{j}\}}=\pi^{-1}(\mathcal{B}(\mathcal{O})_{\{\mathbf{j}\}})$, i.e., the subspace of $\mathcal{S}(\mathcal{O})$ lying over the stratum $\mathcal{B}(\mathcal{O})_{\{\mathbf{j}\}} \subset \mathcal{B}(\mathcal{O})$ of the shifted principal Weyl chamber.
\end{definition}

Theorem~\ref{thr: generic strata} implies that on $\mathcal{S}(\mathcal{O})_{\{\mathbf{j}\}}$ the following conditions hold. For all indices $k \in \overline{\{\mathbf{j}\}} = \{2, \ldots, n\} \setminus \{\mathbf{j}\}$ one has $\psi_k = 0$ and $x_{k - 1} - x_k = c$. Moreover, $\psi_1 \ne 0$, and for $a = 1, \ldots, s$ one has $\psi_{j_a} \ne 0$ and $x_{j_a - 1} - x_{j_a} > c$.

Clearly, the stratification of $\mathcal{B}(\mathcal{O})$ \eqref{eq: Bstat} induce a stratification of $\mathcal{S}(\mathcal{O})$
\begin{equation}\label{eq: Sstat}
    \mathcal{S}(\mathcal{O}) = \bigsqcup_{\{\mathbf{j}\}} \mathcal{S}(\mathcal{O})_{\{\mathbf{j}\}}.
\end{equation}

\subsubsection{The maximal stratum} \label{rem: phase fix}

Let us describe the maximal stratum $\mathcal{S}(\mathcal{O})_{\{2, \ldots, n\}}$.

We will use the following technical fact about the determinant of a Cauchy-type matrix.
\begin{lemma} \label{lem: Cauchy_determinant}
Let $a_1, \ldots, a_n$ and $b_1, \ldots, b_n$ be complex numbers, and let $u_1, \ldots, u_n$ and $v_1, \ldots, v_n$ be complex numbers such that $u_k - v_l \ne 0$ for all $k, l = 1, \ldots, n$. Define the $n \times n$ matrix $C$ with entries
\begin{equation} \label{eq: Cauchy_matrix}
    C_{kl} = \frac{a_k b_l}{u_k - v_l}.
\end{equation}
Then,
\begin{equation} \label{eq: Cauchy_determinant_formula}
    \det(C) = \prod_{k = 1}^n \Big( a_k b_k \Big) \,
        \frac{\displaystyle 
            \prod_{\substack{k, l = 1 \\ k < l}}^n (u_k - u_l) \, 
            \prod_{\substack{k, l = 1 \\ k < l}}^n (v_l - v_k)}
        {\displaystyle \prod_{k, l = 1}^n (u_k - v_l)}.
\end{equation}
\end{lemma}

\begin{theorem}\label{thr: general stratum g}
On the maximal stratum $\mathcal{S}(\mathcal{O})_{\{2, \ldots, n\}}$, the matrix $g$ can be written in the form 
\begin{equation} \label{eq: ndim g}
    g = \alpha \, Y,
\end{equation}
where $\alpha$ is a diagonal unitary matrix,
\begin{equation}
    \alpha = \mathrm{diag}(\alpha_1, \ldots, \alpha_n), \quad |\alpha_k| = 1, \  k = 1, \ldots, n, \text{ and } \prod_{l = 1}^n \alpha_l = 1,
\end{equation}
and $Y$ is a real orthogonal matrix with entries
\begin{equation} \label{eq: X-matrix_maximal_stratum}
    Y_{kl} = \frac{|\lambda_k| |\psi_l|}{x_l - x_k + c},
\end{equation}
with $\det Y  = 1$.
\end{theorem}

\begin{proof}
We start from the momentum map equation \eqref{eq: momentum_map_equation}. Multiplying it from the left by $g$ and using the definition $\lambda = g \psi$, we obtain
\begin{equation}
    g x - x g + c g = \lambda \psi^\dagger.
\end{equation}
In components, this matrix equation gives a system of equations for the elements $g_{kl}$,
\begin{equation}\label{eq: moment map lam psi}
    (x_l - x_k + c) g_{kl} = \lambda_k \, \bar{\psi}_l.
\end{equation}
On the maximal stratum, we have $x_l - x_k + c \ne 0$ for all possible $k, l = 1, \ldots, n$, and therefore
\begin{equation}\label{eq: gelem}
    g_{kl} = \frac{\lambda_k \, \bar{\psi}_l}{x_l - x_k + c}.
\end{equation}

On the maximal stratum, $\psi_l \ne 0$ for all $l = 1, 2, \ldots, n$. The remaining gauge symmetry $U(1)^{\times n} \subset U(n)$ can be used to fix the phases of $\psi_l$, so that all components are real and positive, i.e., $\psi_l = |\psi_l|$ and hence $\bar{\psi}_l = |\psi_l|$ for $l = 1, \ldots, n$. Moreover, on this stratum, one has $\lambda_k \ne 0$ for $k = 1, \ldots, n$. Thus, each $\lambda_k$ can be decomposed as
\begin{equation}
    \lambda_k = \alpha_k |\lambda_k| \, , \quad |\alpha_k| = 1, \quad k = 1, \ldots, n.
\end{equation}
Substituting this to \eqref{eq: gelem}, we obtain the decomposition of $g$ stated in the theorem.

Notice that as $|\alpha_k| = 1$, the diagonal matrix $\alpha$ satisfies $\alpha^\dagger = \alpha^{-1}$. Together with the unitarity of $g$, it implies that the matrix $Y$ is unitary. As all entries of $Y$ are real, it follows that $Y$ is an orthogonal matrix.

The matrix $Y$ \eqref{eq: X-matrix_maximal_stratum} is a matrix of Cauchy type \eqref{eq: Cauchy_matrix} with $a_k = |\lambda_k|, \, b_k = |\psi_k|, \, u_k = c - x_k$ and $v_k = -x_k$. On the maximal stratum, $u_k - v_l = c - x_k + x_l \ne 0$ for all $k, l = 1, \ldots, n$, so the Cauchy determinant formula \eqref{eq: Cauchy_determinant_formula} from Lemma~\ref{lem: Cauchy_determinant} applies and gives
\begin{align} \label{eq: X-determinant-Cauchy}
    \det(Y) &= \prod_{k = 1}^n \Big( |\lambda_k| \, |\psi_k| \Big) \,
        \frac{\displaystyle 
            \prod_{\substack{k, l = 1 \\ k < l}}^n (x_l - x_k) \, 
            \prod_{\substack{k, l = 1 \\ k < l}}^n (x_k - x_l)}
        {\displaystyle \prod_{k, l = 1}^n (c - x_k + x_l)} = \\ &= 
     \prod_{k = 1}^n \Big( |\lambda_k| \, |\psi_k| \Big) \cdot \frac{1}{c^n} 
     \prod_{\substack{k, l = 1 \\ k < l}}^n 
        \frac{x_k - x_l}{x_k - x_l - c} \frac{x_k - x_l}{x_k - x_l + c}.
\end{align}

Using Theorem~\ref{thr: generic strata}(c), we get
\begin{equation} \label{eq: psi-lambda-product}
    \prod_{k = 1}^n \big(|\lambda_k| \, |\psi_k| \big) = c^n
    \prod_{\substack{k, l = 1 \\ k < l}}^n \frac{x_k - x_l - c}{x_k - x_l} \frac{x_k - x_l + c}{x_k - x_l}.
\end{equation}
Combining \eqref{eq: X-determinant-Cauchy} and \eqref{eq: psi-lambda-product}, we get $\det(Y) = 1$. Together with $\det(g) = 1$, it implies that $\det(\alpha) = 1$, i.e., that $\prod_{l = 1}^n \alpha_l = 1$.

\end{proof}

The result of Theorem~\ref{thr: general stratum g} can be summarized as follows.

In the gauge where $x$ is diagonal, the maximal stratum $\mathcal{S}(\mathcal{O})_{\{2, \ldots, n\}} \subset \mathcal{S}(\mathcal{O})$ admits an explicit description: the variables $\psi$ and $g$ are uniquely reconstructed from the eigenvalues of $x$ and the phase variables $\alpha$. 

More precisely, it has a natural coordinate system, consisting of the eigenvalues $(x_1, \ldots, x_n)$ which lie in the interior of the shifted principal Weyl chamber $x_{k - 1} - x_{k} > c,\ k = 2, \ldots, n, \  \sum_{l=1}^n x_{l} = 0$, together with phase variables $(\alpha_1, \ldots, \alpha_n)$, satisfying $|\alpha_k| = 1$, $\ k = 1, \ldots, n$, and $\prod_{l=1}^n \alpha_{l} = 1$. Consequently, $\mathcal{S}(\mathcal{O})_{\{2, \ldots, n\}} \simeq \mathbb{R}_{> 0}^{n - 1} \times \mathbb{T}^{n - 1}$.

\subsubsection{Low-dimensional strata $\mathcal{S}(\mathcal{O})_{\{\mathbf{j}\}}$ } \label{sec: low tori} 

To describe low-dimensional strata, we first establish the following technical fact.

\begin{proposition}\label{prop: zero psi}
Assume $\psi_m = 0$ for some $m = 2, \ldots, n$. Then $g_{m - 1, m}$ is the only nonzero entry of $g$ in the $(m - 1)$-st row and in the $m$-th column. Moreover, $|g_{m - 1, m}| = 1$.
\end{proposition}

\begin{proof} 
Multiplying the momentum map equation \eqref{eq: momentum_map_equation} by $g$ from the left, we obtain
\begin{equation}
    g x - x g + c g = g \psi \psi^\dagger.
\end{equation}
Taking the $(k, m)$-matrix element of this equation gives
\begin{equation} \label{eq: x-g-equation-with-zero}
    (x_m - x_k + c) g_{km} = 0.
\end{equation}

By Theorem~\ref{thr: generic strata}(b), the condition $\psi_m = 0$ is equivalent to $x_{m - 1} - x_m = c$, or equivalently, $x_m - x_{m - 1} + c = 0$. Thus, the coefficient $x_m - x_k + c$ in \eqref{eq: x-g-equation-with-zero} vanishes for $k = m - 1$, so the matrix element $g_{m - 1, m}$ may be nonzero. For all $k \ne m - 1$, one has $x_m - x_k + c \ne 0$, and therefore $g_{km} = 0$. 

If $g_{km} = 0$ for all $k$, then the $m$-th column of $g$ vanishes and $g$ is not invertible, contradicting the fact that $g \in SU(n)$. Hence, the $m$-th column of $g$ has exactly one nonzero entry, namely $g_{m - 1, m}$.

The unitarity condition $g^\dagger g = 1$, applied to the $m$-th column, reads 
\begin{equation}
    \left(g^\dagger g\right)_{km} = 
        \sum_{l = 1}^n \bar{g}_{lk} g_{lm} = 
            \bar{g}_{m - 1, k} g_{m - 1, m} = \delta_{km},
\end{equation}
where we used that $g_{lm} \ne 0$ only for $l = m - 1$. It implies that all entries of $g$ in the $(m - 1)$-st row vanish except for $g_{m - 1, m}$, and moreover $|g_{m - 1, m}| = 1$.
\end{proof}

Fix a subset $\{\mathbf{j}\} = \{j_1, \dots, j_s\} \subseteq \{2, \ldots, n\}$ with $2 \le j_1 < \ldots < j_s \le n$ and set $\overline{\{\mathbf{j}\}} = \{2, \ldots, n\} \backslash \{\mathbf{j}\}$. We now descibe the stratum $\mathcal{S}(\mathcal{O})_{\{\mathbf{j}\}}$.

On the stratum $\mathcal{S}(\mathcal{O})_{\{\mathbf{j}\}}$, one has $\psi_k = 0$ for all $k \in \overline{\{\mathbf{j}\}}$. Hence, by Proposition~\ref{prop: zero psi}, for each $k \in \overline{\{\mathbf{j}\}}$ the matrix $g$ has only one nonzero element $g_{k - 1, k}$ with $|g_{k - 1, k}| = 1$ in the $k$-th column and in the $(k - 1)$-st row.

Define $\hat{g} = P^{-1} g$, where $P$ is the signed matrix of cyclic permutation
\begin{equation}\label{eq: Pmat}
P = 
  \begin{pmatrix}
   0 & -1 & 0 & \ldots &0 \\
   0 & 0 & -1 & \ldots &0 \\
   \vdots & \vdots & \vdots & \ddots & \vdots  \\
   0           & 0 &  0           & \ldots & -1 \\
   1 & 0 & 0 & \ldots & 0 \\
  \end{pmatrix}
.
\end{equation}
Then if $k \in \overline{\{\mathbf{j}\}}$, the diagonal entry $\hat{g}_{kk}$ with $|\hat{g}_{kk}| = 1$ is the only nonzero entry in both the $k$-th column and the $k$-th row of $\hat{g}$. More explicitly, we proved that
\begin{equation}
\begin{aligned}
    |\hat{g}_{kk}| &= 1, \qquad\qquad &&\text{for } k \in \overline{\{\mathbf{j}\}}, \\ 
    \hat{g}_{kl} &= 0, \quad &&\text{for } k, l \in \overline{\{\mathbf{j}\}} \text{ and } k \ne l, \\
    \hat{g}_{kl} &= \hat{g}_{lk} = 0, \quad &&\text{for } k \in \overline{\{\mathbf{j}\}} \text{ and } 
        l \notin \overline{\{\mathbf{j}\}}.
\end{aligned}
\end{equation}

It remains to describe the matrix elements $\hat{g}_{kl}$ for indices $k$ and $l$ such that neither belongs to $\overline{\{\mathbf{j}\}}$.

Define an auxiliary vector $\hat{\lambda} = \hat{g} \psi$ and denote for convenience $j_0 = 1$. Let $x^P = P^{-1} x P$ be the diagonal matrix obtained from $x$ by the cyclic permutation induced by $P$, explicitly, $x^P_k = x_{k - 1}$ for $k = 2, \ldots, n$, and $x^P_1 = x_n$.

\begin{proposition} \label{prop: stratum-g-ja-jb}
On the stratum $\mathcal{S}(\mathcal{O})_{\{\mathbf{j}\}}$, the following statements hold:
\begin{enumerate}

\item[(a)] $\hat{\lambda}_k = 0$ for $k \in \overline{\{\mathbf{j}\}}$.

\item[(b)] For $a, b = 0, 1, \ldots, s$, the matrix elements $\hat{g}_{j_a j_b}$ are given by
\begin{equation}
    \hat{g}_{j_a j_b} = \frac{\hat{\lambda}_{j_a} \bar{\psi}_{j_b}}{x_{j_b} - x^P_{j_a} + c}.
\end{equation}

\item[(c)] For $a = 0, 1, \ldots, s$, the nonvanishing components of $\psi$ and $\hat{\lambda}$ satisfy
\begin{equation}\label{eq: dims psi}
    |\psi_{j_{a}}|^2 = 
        \frac{\displaystyle \prod_{d = 0}^s (x_{j_a} - x^P_{j_d} + c)}
            {\displaystyle \prod_{\substack{d = 0 \\ d \ne a}}^s (x_{j_a} - x_{j_d})}, \quad \  
    |\hat{\lambda}_{j_{a}}|^2 = 
        \frac{\displaystyle \prod_{d = 0}^s (x_{j_d} - x^P_{j_a} + c)}
            {\displaystyle \prod_{\substack{d = 0 \\ d \ne a}}^s (x^P_{j_d} - x^P_{j_a})}. 
\end{equation}

\end{enumerate}
\end{proposition}

\begin{proof}
If $k \in \overline{\{\mathbf{j}\}}$, then we have $\psi_k = 0$. Moreover, as shown above, the matrix $\hat{g}$ has exactly one nonzero entry $\hat{g}_{kk}$ in the $k$-th row. Therefore,
\begin{equation}
    \hat{\lambda}_k = \sum_{l = 1}^n \hat{g}_{kl} \psi_l = \hat{g}_{kk} \psi_k = 0,
\end{equation}
which proves part~(a).

The nonvanishing elements of $\hat{\lambda}$ are $\hat{\lambda}_{j_a}$ for $a = 0, 1, \ldots, s$. From the structure of the matrix $\hat{g}$ proved above, we obtain
\begin{equation}
    \hat{\lambda}_{j_a} = \sum_{d = 0}^s \hat{g}_{j_a j_d} \psi_{j_d}.
\end{equation}

Rewriting the momentum map equation \eqref{eq: momentum_map_equation} in terms of $\hat{g}$ and $x^P$, we obtain
\begin{equation} \label{eq: momentum-hat-g-x-P}
    x - \hat{g}^{-1} x^P \hat{g} = \psi \psi^\dagger - c \cdot 1.
\end{equation}
Multiplying this equation from the left by $\hat{g}$ yields
\begin{equation}
    \hat{g} x - x^P \hat{g} + c \hat{g} = \hat{\lambda} \psi^\dagger.
\end{equation}
In components, this equation gives
\begin{equation} \label{eq: moment map eqs with zeros}
    (x_{j_b} - x^P_{j_a} + c) \hat{g}_{j_a j_b} = \hat{\lambda}_{j_a} \bar{\psi}_{j_b}.
\end{equation}
Since $x_{j_b} - x^P_{j_a} + c \ne 0$ on the stratum $\mathcal{S}(\mathcal{O})_{\{\mathbf{j}\}}$, equation \eqref{eq: moment map eqs with zeros} implies
\begin{equation}
    \hat{g}_{j_a j_b} = \frac{\hat{\lambda}_{j_a} \bar{\psi}_{j_b}}{x_{j_b} - x^P_{j_a} + c},
\end{equation}
which proves part~(b).

Rewrite \eqref{eq: momentum-hat-g-x-P} in the form
\begin{equation}
    \hat{g}^{-1} x^P \hat{g} = x - \psi \psi^\dagger + c \cdot 1.
\end{equation}
Comparing the characteristic polynomials on both sides, we obtain
\begin{equation}
    \prod_{k = 1}^n (x^P_k - z) = \prod_{k = 1}^n (x_k + c - z) - 
        \sum_{b = 0}^s |\psi_{j_b}|^2 \prod_{\substack{k = 1 \\ k \ne j_b}}^n (x_k + c - z),
\end{equation}
where we used Lemma~\ref{lem: rank1 det} to compute the determinant on the right-hand side, together with the fact that $\psi_k = 0$ for $k \in \overline{\{\mathbf{j}\}}$. 

Separating the factors corresponding to indices in $\overline{\{\mathbf{j}\}}$, this identity can be rewritten as
\begin{equation}
    \prod_{d = 0}^s (x^P_{j_d} - z) \prod_{k \in \overline{\{\mathbf{j}\}}} (x^P_k - z) =
        \left(
            \prod_{d = 0}^s (x_{j_d} + c - z) - 
                \sum_{b = 0}^s |\psi_{j_b}|^2 \prod_{\substack{d = 0 \\ d \ne b}}^s (x_{j_d} + c - z)
        \right) \prod_{k \in \overline{\{\mathbf{j}\}}} (x_k + c - z).
\end{equation}
Substituting $z = x_{j_a} + c$, we obtain
\begin{equation}
    \prod_{d = 0}^s (x^P_{j_d} - x_{j_a} - c) 
        \prod_{k \in \overline{\{\mathbf{j}\}}} (x^P_k - x_{j_a} - c) =
    -|\psi_{j_a}|^2 \prod_{\substack{d = 0 \\ d \ne a}}^s (x_{j_d} - x_{j_a})
        \prod_{k \in \overline{\{\mathbf{j}\}}} (x_k - x_{j_a}).
\end{equation}

For $k \in \overline{\{\mathbf{j}\}}$, we have $x^P_k = x_{k - 1} = x_k + c$, and hence $x^P_k - x_{j_a} - c = x_k - x_{j_a}$. Therefore, the products over $k \in \overline{\{\mathbf{j}\}}$ are equal on both sides and cancel. Since the spectrum of $x$ is simple, division by the remaining nonzero factors gives the formula \eqref{eq: dims psi} for $|\psi_{j_a}|^2$.

Similarly, rewriting \eqref{eq: momentum-hat-g-x-P} in the form
\begin{equation}
    \hat{g} \, x \, \hat{g}^{-1} = x^P + \hat{\lambda} \, \hat{\lambda}^\dagger - c \cdot 1
\end{equation}
and applying Lemma \ref{lem: rank1 det} in the same way yields the formula \eqref{eq: dims psi} for $|\hat{\lambda}_{j_a}|^2$.

\end{proof}

\begin{theorem}\label{thr: sdim}
On the stratum $\mathcal{S}(\mathcal{O})_{\{\mathbf{j}\}}$, the matrix $g$ can be written in the form 
\begin{equation}
    g = P \, \hat{g} = P \, \beta \, \hat{Y},
\end{equation}
where $\beta$ is a diagonal unitary matrix,
\begin{equation}
    \beta = \mathrm{diag}(\beta_1, \ldots, \beta_n), \quad 
        \beta_k = 1, \  k \in \overline{\{\mathbf{j}\}}, \quad 
            |\beta_{j_a}| = 1, \ a = 0, 1, \ldots, s, \ \text{ and } \prod_{a = 0}^s \beta_{j_a} = 1,
\end{equation}
$P$ is the signed cyclic permutation matrix \eqref{eq: Pmat}, and $\hat{Y}$ is a real orthogonal matrix with entries
\begin{equation} \label{eq: Y-matrix}
\begin{aligned}
    \hat{Y}_{kl} &= \delta_{kl}, 
          &\text{for }& k, l \in \overline{\{\mathbf{j}\}}, \\
    \hat{Y}_{j_a, k} &= \hat{Y}_{k, j_a} = 0,
           &\text{for }& k \in \overline{\{\mathbf{j}\}}, \text{ and } a = 0, 1, \ldots, s, \\
    \hat{Y}_{j_a, j_b} &= (-1)^{1 - \delta_{a, 0}} 
        \frac{|\hat{\lambda}_{j_a}| |\psi_{j_b}|}{x_{j_b} - x^P_{j_a} + c}, \qquad
          &\text{for }& a, b = 0, 1, \ldots, s.
\end{aligned}
\end{equation}
with $\det \hat{Y} = 1$.
\end{theorem}

\begin{proof}
Proposition~\ref{prop: zero psi} and Proposition~\ref{prop: stratum-g-ja-jb} imply that the matrix $\hat{g} = P^{-1} g$ has the following block structure:
\begin{equation}
\begin{aligned}
    |\hat{g}_{kk}| &= 1, 
        \qquad \qquad \, &&\text{for } k \in \overline{\{\mathbf{j}\}}, \\
    \hat{g}_{kl} &= 0, 
        \qquad \qquad  \, &&\text{for } k, l \in \overline{\{\mathbf{j}\}}, \text{ and } k \ne l, \\
    \hat{g}_{j_a, k} &= \hat{g}_{k, j_a} = 0, 
        \qquad \qquad  \, &&\text{for } k \in \overline{\{\mathbf{j}\}}, \text{ and } a = 0, 1, \ldots, s, \\
    \hat{g}_{j_a, j_b} &= \frac{\hat{\lambda}_{j_a} \bar{\psi}_{j_b}}{x_{j_b} - x^P_{j_a} + c}, 
        \quad &&\text{for } a, b = 0, 1, \ldots, s.
\end{aligned}
\end{equation}

The remaining gauge symmetry $U(1)^{\times n} \subset U(n)$ can be used to set $g_{kk} = 1$ for all $k \in \overline{\{\mathbf{j}\}}$ and to fix the phases of $\psi_{j_a}$, so that $\psi_{j_a} = \bar{\psi}_{j_a} = |\psi_{j_a}|$ for $a = 0, 1, \ldots, s$. After this gauge fixing, we decompose 
\begin{equation}
    \hat{\lambda}_{j_a} = 
        -\beta_{j_a} |\hat{\lambda}_{j_a}|, \quad |\beta_{j_a}| = 1, \quad a = 1, \ldots, s, \quad
        \text{and } \hat{\lambda}_{j_0} = \beta_{j_0} |\hat{\lambda}_{j_0}|.
\end{equation}
The sign convention used here is consistent with that used in the maximal stratum description. Since $\hat{\lambda} = \hat{g} \, \psi = P^{-1} \, g \, \psi$, the variables $\beta_{j_a}$ without $(-1)$ signs are parametrizing the phases of nonzero components of the vector $\lambda = g \, \psi$.

As a result, the matrix $\hat{g}$ has the following form:
\begin{equation}
\begin{aligned}
    \hat{g}_{kl} &= \delta_{kl}, 
        \qquad \qquad   &&\text{for } k, l \in \overline{\{\mathbf{j}\}}, \\
    \hat{g}_{j_a, k} &= \hat{g}_{k, j_a} = 0, 
        \qquad \qquad  \ \ &&\text{for } k \in \overline{\{\mathbf{j}\}}, \text{ and } a = 0, 1, \ldots, s, \\
    \hat{g}_{j_a, j_b} &= (-1)^{1 - \delta_{a, 0}} \beta_{j_a} \, \frac{|\hat{\lambda}_{j_a}| |\psi_{j_b}|}{x_{j_b} - x^P_{j_a} + c}, 
        \quad &&\text{for } a, b = 0, 1, \ldots, s.
\end{aligned}
\end{equation}
Thus, we have shown that the matrix $\hat{g}$ decomposes as $\hat{g} = \beta \, \hat{Y}$, with $\beta$ and $\hat{Y}$ given explicitly as in the statement of the theorem. Consequently, $g = P \, \beta \, \hat{Y}$. 

As in the case of the maximal stratum, the diagonal matrix $\beta$ satisfies $\beta^\dagger = \beta^{-1}$. The matrix $P$ \eqref{eq: Pmat} is a real orthogonal matrix. Therefore, the unitarity of $g$ implies that the matrix $\hat{Y}$ is unitary, and because all its entries are real, the matrix $\hat{Y}$ is orthogonal.

The matrix $\hat{Y}$ has a block structure: the block corresponding to the indices $\overline{\{\mathbf{j}\}}$ is the identity matrix, while the block indexed by $\{j_0, j_1, \ldots, j_s\}$ is a matrix of Cauchy type \eqref{eq: Cauchy_matrix}. Then, we can compute $\det(\hat{Y})$ using the Cauchy determinant formula \eqref{eq: Cauchy_determinant_formula} from Lemma~\ref{lem: Cauchy_determinant}, yielding
\begin{equation}\label{eq: Y-determinant-Cauchy}
    \det(\hat{Y}) = (-1)^{s} \prod_{a = 0}^s \Big( |\hat{\lambda}_{j_a}| \, |\psi_{j_a}| \Big) \,
        \frac{\displaystyle 
            \prod_{\substack{a, b = 0 \\ a < b}}^s (x_{j_b} - x_{j_a}) \, 
            \prod_{\substack{a, b = 0 \\ a < b}}^s (x^P_{j_a} - x^P_{j_b})}
        {\displaystyle \prod_{a, b = 0}^s (x_{j_b} - x^P_{j_a} + c)}, 
\end{equation}
where the sign factor $(-1)^{s}$ comes from $(-1)^{1 - \delta_{a, 0}}$ in the matrix $Y$.

From Proposition~\ref{prop: stratum-g-ja-jb}(c), we obtain
\begin{equation}
    \prod_{a = 0}^s \Big( |\hat{\lambda}_{j_a}|^2 \, |\psi_{j_a}|^2 \Big) = 
        \frac{\displaystyle \prod_{a, b = 0}^s (x_{j_b} - x^P_{j_a} + c)(x_{j_a} - x^P_{j_b} + c)}
            {\displaystyle 
                \prod_{\substack{a, b = 0 \\ a \ne b}}^s (x_{j_a} - x_{j_b}) (x^P_{j_b} - x^P_{j_a})} 
     =  \frac{\displaystyle \prod_{a, b = 0}^s (x_{j_b} - x^P_{j_a} + c)^2}
            {\displaystyle 
                \prod_{\substack{a, b = 0 \\ a < b}}^s (x_{j_a} - x_{j_b})^2 (x^P_{j_b} - x^P_{j_a})^2}.
\end{equation}
Taking the positive square root of this expression, we get
\begin{equation}
    \prod_{a = 0}^s \Big( |\hat{\lambda}_{j_a}| \, |\psi_{j_a}| \Big) = 
        (-1)^{s} \,
        \frac{\displaystyle \prod_{a, b = 0}^s (x_{j_b} - x^P_{j_a} + c)}
        {\displaystyle 
            \prod_{\substack{a, b = 0 \\ a < b}}^s (x_{j_b} - x_{j_a}) \, 
            \prod_{\substack{a, b = 0 \\ a < b}}^s (x^P_{j_a} - x^P_{j_b})}.
\end{equation}

Substituting this expression into \eqref{eq: Y-determinant-Cauchy}, we conclude that $\det(\hat{Y}) = 1$. 

Since $\det(g) = 1$ and $\det(P) = 1$, it follows that $\det(\beta) = 1$, or equivalently
\begin{equation}
    \prod_{a = 0}^s \beta_{j_a} = 1.
\end{equation}

\end{proof}

The result of Theorem~\ref{thr: sdim} can be summarized as follows.

In the gauge where $x$ is diagonal, each stratum $\mathcal{S}(\mathcal{O})_{\{\mathbf{j}\}} \subset \mathcal{S}(\mathcal{O})$ admits an explicit parametrization with all variables are uniquely reconstructed from the eigenvalues of $x$ and the phase variables $\beta$. 

Explicitly, $\mathcal{S}(\mathcal{O})_{\{\mathbf{j}\}}$ carries a natural coordinate system, consisting of the eigenvalues $(x_1, \ldots, x_n)$ satisfying
\begin{equation}
    x_{k - 1} - x_k = c \text{ for } k \in \overline{\{\mathbf{j}\}}, \ 
        x_{j_a - 1} - x_{j_a} > c \text{ for } a = 1, \ldots, s \text{ and } \sum_{l = 1}^n x_l = 0,
\end{equation}
together with phase variables $(\beta_{j_0}, \beta_{j_1}, \ldots, \beta_{j_s})$, satisfying $|\beta_{j_a}| = 1$, $\ a = 0, 1, \ldots, s$, and such that $\prod_{a = 0}^s \beta_{j_a} = 1$. 

Consequently, for $s > 0$, each stratum $\mathcal{S}(\mathcal{O})_{\{\mathbf{j}\}} \simeq \mathbb{R}_{> 0}^s \times \mathbb{T}^s$.

\begin{remark} \label{rem: zero-dimensional-stratum}
If $s = 0$, the stratum $\mathcal{S}(\mathcal{O})_\varnothing$ is zero-dimensional, and it contains only one point.

In this case, the eigenvalues of $x$ are fixed by the constraints $x_{k - 1} - x_k = c$ for $k = 2, \ldots, n$ and $\sum_{l = 1}^n x_l = 0$, which uniquely determine
\begin{equation}
    x_k = \frac{n - 1}{2} \,c - (k - 1) c, \quad k = 1, \ldots, n.
\end{equation}

The phase-variable matrix is trivial: $\beta = 1$.

The only nonvanishing components of $\psi$ and $\hat{\lambda}$ are $\psi_1$ and $\hat{\lambda}_1$, respectively, and they satisfy $|\psi_1|^2 = nc$ and $|\hat{\lambda}_1|^2 = nc$. It implies that the matrix $\hat{Y}$ is the identity: the only potentially nontrivial entry is
\begin{equation}
    \hat{Y}_{11} = \frac{|\hat{\lambda}_1| |\psi_1|}{x_1 - x^P_1 + c} = 
        \frac{\sqrt{nc} \cdot \sqrt{nc}}{x_1 - x_n + c} = \frac{nc}{(n - 1)c + c} = 1,
\end{equation}
while $\hat{Y}_{22} = \ldots = \hat{Y}_{nn} = 1$, and $\hat{Y}_{kl} = 0$ for $k, l = 1, \ldots, n$ and $k \ne l$. Thus, $\hat{Y} = 1$ on the zero-dimensional stratum, and consequently, $g = P$. 

The eigenvalues of $g$ are $-1, -\omega, \ldots, -\omega^{n - 1}$, where $\omega = e^{2 \pi i /n}$ is a primitive $n$-th root of unity. In terms of the original CMS dynamics, these eigenvalues correspond to the particle coordinates of the CMS particles $q_k = \frac{2 \pi k}{n} - \pi$ at the unique fixed point of the multi-time CMS dynamic, up to $S_n$ permutations of coordinates.
\end{remark}

\begin{remark} \label{rem: two-parametrizations-alpha-beta}
Notice that on the stratum $\mathcal{S}(\mathcal{O})_{\{\mathbf{j}\}}$ the matrix $g$ can be equivalently written in the form 
\begin{equation} \label{eq: ndim g}
    g = \alpha \, P \, \hat{Y},
\end{equation}
where $\alpha$ is a diagonal unitary matrix related to $\beta$ by $\alpha = P \, \beta \, P^{-1}$. In terms of components, the nontrivial phase variables are $\alpha_{j_a - 1} = \beta_{j_a}$ for $a = 1, \ldots, s$, and $\alpha_n = \beta_1$. The constraint $\prod_{a = 0}^s \beta_{j_a} = 1$, written in terms of $\alpha$-variables, takes the form
\begin{equation}
    \alpha_n \cdot \prod_{a = 1}^s \alpha_{j_a - 1} = 1.
\end{equation}

This representation is compatible with the parametrization of the maximal stratum $\mathcal{S}(\mathcal{O})_{\{2, \ldots, n\}}$, where $g$ admits the decomposition $g = \alpha Y$ with diagonal unitary $\alpha$ and real orthogonal $Y$. 

This parametrization will be used throughout the rest of the paper.
\end{remark}

\subsection{The dynamical stratification of the phase space} \label{sec: walls}

It is straightforward to verify that the closure of each stratum $\mathcal{B}(\mathcal{O})_{\{\mathbf{j}\}}$ of the base $\mathcal{B}(\mathcal{O})$ contains all strata lying on its boundary. Equivalently, the closure of $\mathcal{B}(\mathcal{O})_{\{\mathbf{j}\}}$ contains all strata labeled by subsets of $\{\mathbf{j}\}$ 
\begin{equation}
    \overline{\mathcal{B}(\mathcal{O})_{\{\mathbf{j}\}}}=
        \bigsqcup_{\{\mathbf{j}'\}\subseteq\{\mathbf{j}\}} \mathcal{B}(\mathcal{O})_{\{\mathbf{j}'\}}.
\end{equation}

The same stratification property holds for the reduced phase space,
\begin{equation}
    \overline{\mathcal{S}(\mathcal{O})_{\{\mathbf{j}\}}} = 
        \bigsqcup_{\{\mathbf{j}' \}\subseteq \{\mathbf{j}\}} \mathcal{S}(\mathcal{O})_{\{\mathbf{j}'\}}.
\end{equation}

This structure can be verified explicitly by considering appropriate degeneration limits. On the stratum $\mathcal{S}(\mathcal{O})_{\{\mathbf{j}\}}$, the eigenvalues of $x$ satisfy $x_{j_a - 1} > x_{j_a} + c$ for $a = 1, \ldots, s$. Fix an index $b \in \{1, \ldots, s\}$ and consider the limit $x_{j_b - 1} \to x_{j_b} + c$.

Form the explicit expressions \eqref{eq: dims psi}, we see that in this limit the quantities $|\psi_{j_b}|^2$ and $|\hat{\lambda}_{j_b}|^2$ tend to zero,
\begin{equation}
    |\psi_{j_b}|^2 \to 0, \quad |\hat{\lambda}_{j_b}|^2 \to 0,
\end{equation}
since their numerators contain the factor $x_{j_b} - x^P_{j_b} + c \to 0$. As $|\hat{\lambda}_{j_b}|^2 \to 0$, its phase can not define an angular variable in the limit, and the dimension of the torus of phase variables reduces from $s$ to $s - 1$.

At the same time, for $a \ne b$ the quantities $|\psi_{j_a}|^2$ and $|\hat{\lambda}_{j_a}|^2$ remain strictly positive. In the limit, the products appearing in \eqref{eq: dims psi} lose the factors corresponding to $d = b$, due to the identification $x^P_{j_b} \to x_{j_b} + c$.

As a result, the limiting expressions coincide precisely with the formulas describing the stratum $\mathcal{S}(\mathcal{O})_{\{\mathbf{j}\} \setminus \{j_b\}}$. By iterating this procedure, we can get all strata $\mathcal{S}(\mathcal{O})_{\{\mathbf{j}'\}}$ corresponding to arbitrary subsets $\{\mathbf{j}'\} \subset \{\mathbf{j}\}$, confirming the claimed closure relations.

\section{Action-angle variables on $\mathcal{S}(\mathcal{O})_{\{\mathbf{j}\}}$} \label{sec: action-angle-variables-on-strata}

\subsection{Natural coordinates on $\mathcal{S}(\mathcal{O})_{\{\mathbf{j}\}}$} \label{ytheta}

In this section, we describe convenient coordinate systems on the strata $\mathcal{S}(\mathcal{O})_{\{\mathbf{j}\}}$

\subsubsection{The maximal stratum}
On the maximal stratum $\mathcal{S}(\mathcal{O})_{\{2, \ldots, n\}}$, introduce variables 
\begin{equation}
    y_k = x_{k - 1} - x_k, \quad k = 2, \ldots, n,
\end{equation}
which satisfy $y_k > c$. In terms of these variables, the eigenvalues of $x$ can be reconstructed as
\begin{equation}\label{eq: x to y}
    x_k = \sum_{l = 2}^n \frac{n + 1 - l}{n} \, y_l - \sum_{l = 2}^k y_l, \quad
        k = 1, \ldots, n.
\end{equation}

Next, write the phase variables as $\alpha_k = e^{i \eta_k}$ for $k = 1, \ldots, n$, where the variables $\eta_k$ are understood modulo $2 \pi$. The constraint $\prod_{l = 1}^n \alpha_l = 1$ implies $\sum_{l = 1}^n \eta_l = 0 \mod 2 \pi$. Define
\begin{equation} \label{eq: theta-angles-maximal-stratum}
    \theta_k = \sum_{l = 1}^{k - 1} \eta_l \mod 2 \pi, \quad k = 2, \ldots, n.
\end{equation}

Together, these variables provide a convenient coordinate system on the maximal stratum 
\begin{equation}
    \mathcal{S}(\mathcal{O})_{\{2, \ldots, n\}} \simeq \mathbb{R}_{> 0}^{n - 1} \times \mathbb{T}^{n - 1}.
\end{equation}
The real coordinates $(y_2, \ldots, y_{n})$ with constraints $y_k > c$ parametrize $\mathbb{R}_{> 0}^{n - 1}$, while the variables $(\theta_2, \ldots, \theta_{n})$ defined modulo $2 \pi$ parametrize the torus $\mathbb{T}^{n - 1}$. In Section \ref{sec: Action-angle variables}, we will compute the symplectic form in coordinates $(y, \theta)$ and show that they are action-angle variables on the maximal stratum.

\subsubsection{Low-dimensional strata}
On a stratum $\mathcal{S}(\mathcal{O})_{\{\mathbf{j}\}}$, introduce the variables 
\begin{equation}
    y_{j_a} = x_{j_a - 1} - x_{j_a}, \quad a = 1, \ldots, s,
\end{equation}
which satisfy $y_{j_a} > c$. In terms of these variables, the eigenvalues of $x$ can be expressed as
\begin{equation}\label{eq: sdim x to y}
    x_k = \frac{n - 1}{2} c - (k - 1)c +
    \sum_{a = 1}^s \left( \frac{n + 1 - j_a}{n} - \delta_{k \ge j_a} \right) (y_{j_a} - c),
\end{equation}
where $\delta_{k \ge l}$ denotes the step function defined by
\begin{equation}
    \delta_{k \ge l} =
        \begin{cases}
            1, & k \ge l, \\
            0, & k < l.
        \end{cases}
\end{equation}

Write the nontrivial phase variables (cf.~Remark~\ref{rem: two-parametrizations-alpha-beta}) as $\alpha_{j_a - 1} = e^{i \eta_{j_a - 1}}$ for $a = 1, \ldots, s$ and $\alpha_n = e^{i \eta_n}$, where the angular variables are understood modulo $2 \pi$. The constraint $\alpha_n \cdot \prod_{a = 1}^s \alpha_{j_a - 1} = 1$ implies $\eta_n + \sum_{a = 1}^s \eta_{j_a - 1} = 0 \mod 2 \pi$. Define
\begin{equation} \label{eq: theta-angles-low-dim-strata}
    \theta_{j_a} = \sum_{b = 1}^a \eta_{j_b - 1} \mod 2 \pi, \quad a = 1, \ldots, s.
\end{equation}
Notice that this definition is consistent with the one given for the maximal stratum, if we set $s = n - 1$ and $\{\mathbf{j}\} = \{2, \ldots, n\}$.

The variables $(y_{j_a}, \theta_{j_a})$ with $y_{j_a} > c$ provide a coordinate system on $\mathcal{S}(\mathcal{O})_{\{\mathbf{j}\}}$. We will show that these coordinates form a system of action-angle variables on this stratum.

\subsubsection{Example: a two-dimensional stratum} \label{sec: two-dimensional-stratum}

We illustrate the general construction in the simplest nontrivial case $s = 1$. Fix an index $j \in \{2, \ldots, n\}$, and consider the corresponding subset $\{\mathbf{j}\} = \{j\}$.

On the stratum $\mathcal{S}(\mathcal{O})_{\{j\}}$, the eigenvalues of the matrix $x$ satisfy $x_{k - 1} - x_{k} = c$ for $k = 2, \ldots, n$ and $k \ne j$, while the remaining difference $y = x_{j - 1} - x_j$ is strictly greater than $c$. Explicitly,
\begin{equation}
    x_{k} = \frac{n - 1}{2} c - (k-1)c +
        \left( \frac{n + 1 - j}{n} - \delta_{k \ge j} \right) (y - c).
\end{equation}

The only nonzero elements of the vector $\psi$ are $\psi_1$ and $\psi_j$. Proposition \ref{prop: stratum-g-ja-jb} implies that the vector $\hat{\lambda}$ also has only two nonzero components, namely $\hat{\lambda}_1$ and $\hat{\lambda}_j$, and that
\begin{equation}
\begin{aligned}  
    &|\psi_{1}|^2 = (j - 1) c \, \frac{y + (n - 1) c}{y + (j - 2) c}, \qquad  \ \quad
    &&|\psi_{j}|^2 = (n - j + 1) c \, \frac{y - c}{y + (j - 2) c}, \\
    &|\hat{\lambda}_{1}|^{2} = (n - j + 1) c \frac{y + (n - 1) c}{y + (n - j) c}, \qquad
    &&|\hat{\lambda}_{j}|^{2} = (j - 1) c \ \frac{y-c}{y + (n - j) c}.
\end{aligned}
\end{equation}

We parametrize the matrix $g$ as $g = \alpha \, P \, \hat{Y}$ as described in Remark~\ref{rem: two-parametrizations-alpha-beta}. The diagonal unitary matrix $\alpha$ has the form
\begin{equation}
    \alpha = \mathrm{diag}\left(1, \ldots, 1, \alpha_{j - 1} = e^{i \theta}, 1, \ldots, 1, \alpha_n = e^{-i \theta}\right),
\end{equation}
$P$ is the signed permutation matrix \eqref{eq: Pmat}, and the real orthogonal matrix $\hat{Y}$ is determined by the general formula \eqref{eq: Y-matrix} from Theorem~\ref{thr: sdim}
\begin{equation}\label{eq: g 2dim}
    \hat{Y} =
    \begin{pmatrix}
        R & 0 &\ldots & Q & \ldots & 0 \\
        0 & 1 &\ldots & 0 & \ldots & 0 \\
        \vdots & \vdots & \ddots & \vdots & \ddots & \vdots  \\
        -Q & 0 & \ldots & R & \ldots & 0 \\
        \vdots & \vdots & \ddots & \vdots & \ddots & \vdots  \\
        0 & 0 & \ldots & 0 & \ldots & 1 \\
    \end{pmatrix},
\end{equation}
where the coefficients $R$ and $Q$ are given by
\begin{equation}\label{eq: rq}
    R^2  = \frac{(j - 1) c \, (n - j + 1) c}{(y + (j - 2) c)(y + (n - j) c)}, \qquad
    Q^2  = \frac{(y - c) (y + (n - 1) c)}{(y + (j - 2) c)(y + (n - j) c)}.
\end{equation}
Note that $R^2 + Q^2 = 1$, and $\hat{Y}$ is orthogonal.

As a result, we have a parametrization of the matrix $g$
\begin{equation}\label{eq: g-matrix-minimal-stratum}
    g =
\begin{pmatrix}
   0 & -1 & 0  & \ldots & 0 & \ldots &0 \\
   0 &  0 & -1 & \ldots & 0 & \ldots &0 \\
   \vdots & \vdots & \vdots & \ddots & \vdots & \ddots & \vdots  \\
   Q e^{i\theta} & 0 & 0 & \ldots & -R e^{i\theta} &\ldots &0 \\
   \vdots & \vdots & \vdots & \ddots & \vdots & \ddots & \vdots \\
   0      & 0      &  0           & \ldots & 0 & \ldots &-1\\
   R e^{-i\theta}  & 0 & 0 & \ldots & Q e^{-i\theta} & \ldots &0\\
\end{pmatrix}.
\end{equation}

Thus, the stratum $\mathcal{S}(\mathcal{O})_{\{j\}}$ is explicitly parametrized by a single action–angle pair $(y, \theta)$, in complete agreement with the general description of low-dimensional strata.
 
\subsection{Symplectic form on $\mathcal{S}(\mathcal{O})_{\{\mathbf{j}\}}$}
\label{sec: Action-angle variables}

Since the variables $\{y\}$ parametrize the base $\mathcal{B}(\mathcal{O})_{\{\mathbf{j}\}}$ of the Lagrangian fibration (\ref{LP}) of $\mathcal{S}(\mathcal{O})_{\{\mathbf{j}\}}$, they can be taken as action variables. In this section, we show that the variables $\{\theta\}$ are the corresponding angle variables. This is achieved by describing the symplectic form on these strata in terms of the coordinates $(y, \theta)$.

\subsubsection{The maximal stratum}

\begin{proposition}\label{prop: ndim Omega}
On the maximal stratum $\mathcal{S}(\mathcal{O})_{\{2, \ldots, n\}}$ the symplectic form is given by
\begin{equation}
    \Omega_{\mathcal{S}(\mathcal{O})_{\{2, \ldots, n\}}} = 
        - \sum_{k = 2}^n \mathrm{d} y_k \wedge \mathrm{d} \theta_k.
\end{equation}
\end{proposition}

\begin{proof}
Using the $SU(n)$-invariance of the symplectic form $\Omega$ \eqref{eq: symplectic_form_T^*G} on $T^* SU(n)$, we evaluate it on the representative $(x, g)$ with the diagonal $x$ and $g = \alpha Y$ as in \eqref{eq: ndim g}. We obtain
\begin{equation}
    \Omega_{\mathcal{S}(\mathcal{O})_{\{2, \ldots, n\}}} = 
    i \, \mathrm{d} \mathrm{Tr} \left( x\, \mathrm{d} g\, g^{-1} \right) =
        i\, \mathrm{d} \mathrm{Tr} \left( x\, \mathrm{d} (\alpha Y)\, Y^{-1} \alpha^{-1} \right) = 
            i\, \mathrm{d} \mathrm{Tr} \left( x\, \mathrm{d} Y\, Y^{-1} \right) +
                i\, \mathrm{d} \mathrm{Tr} \left( x\, \mathrm{d} \alpha\, \alpha^{-1} \right).
\end{equation}

The first term vanishes. Indeed, since $Y$ is orthogonal, the matrix $\mathrm{d} Y \, Y^{-1}$ is skew-symmetric, while $x$ is diagonal and hence symmetric, so their product has zero trace. Equivalently, in components,
\begin{equation} \label{eq: OO zero}
    \mathrm{Tr} \left( x\, \mathrm{d} Y\, Y^{-1} \right) =
        \sum_{k, l = 1}^n x_{k} \, \mathrm{d} Y_{kl} \cdot (Y^{t})_{lk} =
            \frac{1}{2} \sum_{k, l = 1}^n x_{k} \, \mathrm{d} (Y_{kl})^{2} = 
                \frac{1}{2} \sum_{k = 1}^n x_{k}\, \mathrm{d} \sum_{l = 1}^n (Y_{kl})^{2} = 0,
\end{equation}
where we used the orthogonality condition $\sum_{l = 1}^n (Y_{kl})^{2} = 1$. 

Therefore, on the maximal stratum, we obtain
\begin{equation}
    \Omega_{\mathcal{S}(\mathcal{O})_{\{2, \ldots, n\}}} = 
        i \, \mathrm{d} \mathrm{Tr} \left( x\, \mathrm{d} \alpha\, \alpha^{-1} \right) =
            i \sum_{k = 1}^n \mathrm{d} x_{k} \wedge \alpha^{-1}_{k} \mathrm{d}\alpha_{k} =
                - \sum_{k = 1}^n \mathrm{d} x_{k} \wedge \mathrm{d} \eta_{k},
\end{equation}
where we used the parametrization $\alpha_k = e^{i \eta_k}$.

Using the constraint $\sum_{l = 1}^n \eta_l = 0 \mod 2 \pi$, we obtain for the differentials
\begin{equation}
    \mathrm{d} \eta_n = - \sum_{k = 1}^{n - 1} \mathrm{d} \eta_k.
\end{equation}
Using this relation, we can eliminate $\mathrm{d} \eta_n$. The symplectic form can be rewritten as
\begin{equation}
    \Omega_{\mathcal{S}(\mathcal{O})_{\{2, \ldots, n\}}} = 
        - \sum_{k = 1}^{n - 1} \mathrm{d} (x_k - x_n) \wedge \mathrm{d} \eta_k.
\end{equation}
Substituting $x_k - x_n = \sum_{l = k + 1}^n y_l$, we obtain
\begin{equation}
    \Omega_{\mathcal{S}(\mathcal{O})_{\{2, \ldots, n\}}} = 
        - \sum_{k = 1}^{n - 1} \left(
            \sum_{l = k + 1}^n \mathrm{d} y_l
        \right) \wedge \mathrm{d} \eta_k =
        - \sum_{k = 2}^n \mathrm{d} y_k \wedge \left(
            \sum_{l = 1}^{k - 1} \mathrm{d} \eta_l
        \right),
\end{equation}
where in the last step we exchanged the order of summation. By the definition of $\theta_k$ \eqref{eq: theta-angles-maximal-stratum}, this gives
\begin{equation}
    \Omega_{\mathcal{S}(\mathcal{O})_{\{2, \ldots, n\}}} = 
        - \sum_{k = 2}^n \mathrm{d} y_k \wedge \mathrm{d} \theta_k.
\end{equation}
\end{proof}

Thus, on $\mathcal{S}(\mathcal{O})_{\{2, \ldots, n\}}$ the variables $(y_k, \theta_k)$, $k = 2, \ldots, n$, are Darboux coordinates, and $(y_k)$ and $(\theta_k)$ are action and angle variables, respectively.

\subsubsection{Low-dimensional strata}

\begin{proposition}\label{prop: sdim Omega}
On the $2s$-dimensional stratum $\mathcal{S}(\mathcal{O})_{\{\mathbf{j}\}}$, the symplectic form is given by
\begin{equation}\label{eq: sdim Omega}
    \Omega_{\mathcal{S}(\mathcal{O})_{\{\mathbf{j}\}}} = 
        - \sum_{a = 1}^s \mathrm{d} y_{j_{a}} \wedge \mathrm{d}\theta_{j_{a}}.
\end{equation}
\end{proposition}

\begin{proof}
We evaluate the symplectic form on the representative with the diagonal $x$ and $g = \alpha \, P \, \hat{Y}$ as in Remark~\ref{rem: two-parametrizations-alpha-beta}. Then, since $P$ is constant, we obtain
\begin{multline}\notag
    \Omega_{\mathcal{S}(\mathcal{O})_{\{\mathbf{j}\}}} = 
        i\, \mathrm{d} \mathrm{Tr} \left( x\, \mathrm{d} g\, g^{-1} \right) =
            i\, \mathrm{d} \mathrm{Tr} \left( 
                x \, \mathrm{d} (\alpha P \hat{Y})\, \hat{Y}^{-1} P^{-1} \alpha^{-1}\right) = \\ =
        i\, \mathrm{d} \mathrm{Tr} \left( x \, \mathrm{d} \alpha\, \alpha^{-1}\right) +
            i\, \mathrm{d} \mathrm{Tr} \left( x^P \, \mathrm{d} \hat{Y}\, \hat{Y}^{-1}\right),
\end{multline}  
where $x^{P} = P^{-1} x P$.

The term $\mathrm{Tr} \left( x^P \, \mathrm{d} \hat{Y}\, \hat{Y}^{-1}\right) = 0$ vanishes for the same reasons as in equation \eqref{eq: OO zero}. Therefore,
\begin{multline}\notag
    \Omega_{\mathcal{S}(\mathcal{O})_{\{\mathbf{j}\}}} = 
        i\, \mathrm{d} \mathrm{Tr} \left( x \, \mathrm{d} \alpha\, \alpha^{-1}\right) = 
        i\, \mathrm{d} x_n \wedge \alpha_n^{-1}\, \mathrm{d} \alpha_n + 
        i\, \sum_{a = 1}^s \mathrm{d} x_{j_a - 1} \wedge 
            \alpha_{j_a - 1}^{-1} \, \mathrm{d} \alpha_{j_a - 1} = \\ =
        - \mathrm{d} x_n \wedge \mathrm{d} \eta_n -
            \sum_{a = 1}^s \mathrm{d} x_{j_a - 1} \wedge \mathrm{d} \eta_{j_a - 1}.
\end{multline} 

The constraint $\alpha_n \cdot \prod_{b = 1}^s \alpha_{j_b - 1} = 1$ implies $\mathrm{d} \eta_n = - \sum_{b = 1}^s \mathrm{d} \eta_{j_b - 1}$. Eliminating $\mathrm{d} \eta_n$, we obtain
\begin{equation} \label{eq: omega_on_orbit}
    \Omega_{\mathcal{S}(\mathcal{O})_{\{\mathbf{j}\}}} = 
        - \sum_{a = 1}^s \mathrm{d} (x_{j_a - 1} - x_n) \wedge \mathrm{d} \eta_{j_a - 1}.
\end{equation}

From \eqref{eq: sdim x to y}, we have $\mathrm{d} (x_{j_a - 1} - x_n) = \sum_{b = a}^s \mathrm{d} y_{j_b}$. Substituting this into \eqref{eq: omega_on_orbit}, we obtain
\begin{equation}
\begin{aligned}
    \Omega_{\mathcal{S}(\mathcal{O})_{\{\mathbf{j}\}}} = 
        - \sum_{a = 1}^s \left(
            \sum_{b = a}^s \mathrm{d} y_{j_b} 
        \right) \wedge \mathrm{d} \eta_{j_a - 1} =
        - \sum_{a = 1}^s \mathrm{d} y_{j_a} \wedge \left( 
            \sum_{b = 1}^a \mathrm{d} \eta_{j_b - 1} \right).
\end{aligned}
\end{equation}

In terms of the angle $\theta_{j_a} = \sum_{b = 1}^a \eta_{j_b - 1} \mod 2 \pi$ \eqref{eq: theta-angles-low-dim-strata}, we have $\mathrm{d} \theta_{j_a} = \sum_{b = 1}^a \mathrm{d} \eta_{j_b - 1}$, and hence \eqref{eq: sdim Omega}.

\end{proof}

Thus, the coordinates $(y_{j_a}, \theta_{j_a})$ form as system of Darboux coordinates on $\mathcal{S}(\mathcal{O})_{\{\mathbf{j}\}}$. The variables $y_{j_a}$ provide coordinates on the corresponding stratum $\mathcal{B}(\mathcal{O})_{\{\mathbf{j}\}}$ of $\mathcal{B}(\mathcal{O})$. The complementary coordinates $\theta_{j_a}$ parametrize the Lagrangian fibers $\pi^{-1}(y)$, which are tori, and therefore are angle coordinates. 

\subsection{The multi-time dynamics of Calogero--Moser--Sutherland system} \label{sec: multi-time-dynamics}

We now show that the multi-time CMS dynamics is linear in the angle variables $\theta$.

\subsubsection{The multi-time dynamics on $T^* SU(n)$}
\label{sec: The evolution before reduction.}

Let $F$ be a pull-back of a polynomial $SU(n)$-invariant function on $\mathfrak{su}(n)^*$ to $T^* SU(n) \simeq \mathfrak{su}(n)^* \times SU(n)$ with respect to the projection onto the first factor. Thus, $F(x, g) = F(x)$ is a symmetric polynomial in $x_1, \ldots, x_n$, or, equivalently, a polynomial in the invariants $\mathcal{H}_k(x) = \frac{1}{k} \mathrm{Tr}(x^k)$ \eqref{eq: CMHams} for $k = 2, \ldots, n$.

Denote by $v_F$ the Hamiltonian vector field generated by $F$. In components, this vector field is given by
\begin{equation}
    v_F(x, g) = -i \sum_{k, l = 1}^n (\nabla F(x)  \, g)_{kl} \, \frac{\partial}{\partial g_{kl}}.
\end{equation}
Here $\nabla F(x)$ denotes the gradient of $F$ defined by $\mathrm{d} F(x) = \mathrm{Tr}(\nabla F(x) \mathrm{d} x)$. The integral curve of $v_F$ passing through $(x, g)$ at $t = 0$ is given by
\begin{equation}
    (x(t), g(t)) = (x, e^{-i t \nabla F(x)}g).
\end{equation}

Any two such invariant functions $F_1, F_2$  Poisson commute with respect to the standard Poisson bracket on $T^* SU(n)$, $\{F_1, F_2\} = 0$. Consequently, the corresponding Hamiltonian vector fields commute $[v_{F_1}, v_{F_2}] = 0$, and the associated Hamiltonian flows are compatible.

As a result, a multi-time Calogero--Moser--Sutherland dynamics generated by the commuting Hamiltonians $\mathcal{H}_2, \ldots, \mathcal{H}_n$ is well-defined with trajectories given by
\begin{equation}
    (x(\mathbf{t}), g(\mathbf{t}))=(x, e^{- i \sum_{k = 2}^n t_k \nabla \mathcal{H}_k(x)} g),
\end{equation}
where $\mathbf{t} = (t_2, \dots, t_n)$ are times corresponding to the Hamiltonians $\mathcal{H}_2, \ldots, \mathcal{H}_n$.

This dynamics preserves the momentum map,
\begin{equation}
    \mu(x(\mathbf{t}), g(\mathbf{t})) = \mu(x, g) = i(x - g^{-1} x g).
\end{equation}
Consequently, the commuting Hamiltonian flows on $T^* SU(n)$ induce a compatible multi-time dynamics on the reduced phase space $\mathcal{S}(\mathcal{O})$.

\subsubsection{The multi-time dynamics on $\mathcal{S}(\mathcal{O})_{\{\mathbf{j}\}}$}
\label{sec: The multi-time dynamics for s-tori strata.}

We describe now the CMS dynamics on each stratum $\mathcal{S}(\mathcal{O})_{\{\mathbf{j}\}}$. The symplectic form on $\mathcal{S}(\mathcal{O})_{\{\mathbf{j}\}}$ in coordinates $(y_{j_a}, \theta_{j_a})$ is given by
\begin{equation}\label{eq: su(n) form}
    \Omega_{\mathcal{S}(\mathcal{O})_{\{\mathbf{j}\}}} = - \sum_{a = 1}^{s} \mathrm{d}y_{j_{a}} \wedge \mathrm{d}\theta_{j_{a}}.
\end{equation}
This symplectic form implies
\begin{equation}
    \{ y_{j_{a}}, y_{j_{b}}\} = 0, \qquad 
    \{\theta_{j_{a}}, \theta_{j_{b}}\} = 0, \qquad 
    \{\theta_{j_{a}}, y_{j_{b}}\} = \delta_{ab}.
\end{equation}

Since the CMS Hamiltonians $\mathcal{H}_k$ depend only on $x$, the corresponding Hamiltonian vector fields $v_{\mathcal{H}_k}$, when restricted to $\mathcal{S}(\mathcal{O})_{\{\mathbf{j}\}}$, take the form 
\begin{equation}
    v_{\mathcal{H}_{k}}^{\{\mathbf{j}\}} = 
        - \sum_{l = 1}^n \sum_{a = 1}^s 
            \frac{\partial \mathcal{H}_k}{\partial x_l} \frac{\partial x_l}{\partial y_{j_a}} 
                \frac{\partial}{\partial \theta_{j_a}}.
\end{equation}

Using the expression \eqref{eq: sdim x to y}, we can express the eigenvalues $x_l$ in terms of the variables $y_{j_a}$. This yields
\begin{equation}
    v_{\mathcal{H}_k}^{\{\mathbf{j}\}} = - \sum_{a = 1}^s u^{\mathcal{H}_k, \{\mathbf{j}\}}_a(y) \frac{\partial}{\partial \theta_{j_a}},
\end{equation}
where 
\begin{equation}
    u^{\mathcal{H}_k, \{\mathbf{j}\}}_a(y) = 
        \frac{n + 1 - j_a}{n} \sum_{l = 1}^{n} \frac{\partial \mathcal{H}_k}{\partial x_l}(x(y)) - 
            \sum_{l = j_a}^{n} \frac{\partial \mathcal{H}_k}{\partial x_l}(x(y)) .
\end{equation}
Here, the function $x(y)$ is given by (\ref{eq: sdim x to y}).

As a consequence, the multi-time dynamics in $(y, \theta)$ coordinates is given by
\begin{equation} \label{eq: multi-time-dynamics-for-angle}
    y_{j_a}(\mathbf{t}) = y_{j_a}, \qquad 
    \theta_{j_a}(\mathbf{t}) = \theta_{j_a} - \sum_{k = 2}^n u^{\mathcal{H}_k, \{\mathbf{j}\}}_a(y) \, t_k,
        \qquad a = 1, \ldots, s,
\end{equation}
with $\mathbf{t} = (t_2, \dots, t_n)$. 

Thus, the multi-time dynamics is linear in angle variables, as expected for action-angle coordinates. Moreover, although the Hamiltonians $\mathcal{H}_2, \ldots, \mathcal{H}_n$ are independent on $T^* SU(n)$, their restrictions to $\mathcal{S}(\mathcal{O})_{\{\mathbf{j}\}}$ become functionally dependent, since the dimension of the Liouville tori is $s < n - 1$.

\subsubsection{Example: a two-dimensional stratum} \label{sec: motion-on-two-dim-stratum}

We now illustrate the multi-time dynamics on a two-dimensional stratum described in Section~\ref{sec: two-dimensional-stratum} in terms of the original CMS coordinates. On this stratum, there is a single angle variable $\theta$, and the multi-time evolution \eqref{eq: multi-time-dynamics-for-angle} takes the form
\begin{equation}
    y(\mathbf{t}) = y, \qquad
    \theta(\mathbf{t}) = \theta - \sum_{k = 2}^n u^{\mathcal{H}_k, \{j\}}(y) \, t_k.
\end{equation}
In this case, all commuting Hamiltonians $\mathcal{H}_2, \ldots, \mathcal{H}_n$ generate shifts of the same angle variable $\theta$. The corresponding Hamiltonian vector fields are all parallel, so the $(n - 1)$ commuting Hamiltonians become maximally dependent when restricted to this two-dimensional stratum.

Using the explicit expression for the matrix $g$ in terms of the variables $y$ and $\theta$ given in \eqref{eq: g-matrix-minimal-stratum}, we compute its characteristic polynomial
\begin{equation}\label{eq: 1dim spec curv}
    \mathrm{det} \left(z \cdot 1 - g \right) = 
    z^{n} - (-1)^{j} \, Q \, e^{i \theta(\mathbf{t})} z^{n - j + 1} - 
        (-1)^{n - j} \, Q \, e^{-i \theta(\mathbf{t})} z^{j - 1} + (-1)^{n}.
\end{equation}

The zeros of the polynomial \eqref{eq: 1dim spec curv} describe the evolution of the CMS particle coordinates. More precisely,
\begin{equation}
    \mathrm{det} \left(z \cdot 1 - g \right) = \prod_{k = 1}^n (z - \gamma_k(\mathbf{t})) =
        \prod_{k = 1}^n (z - e^{i q_k(\mathbf{t})}).
\end{equation}

\end{document}